\documentclass[aps,prd,reprint,groupedaddress]{revtex4-2}
\usepackage{graphicx}
\usepackage{amsmath}
\usepackage{amssymb}
\usepackage{amsfonts}
\usepackage{dcolumn}
\usepackage{bm}
\usepackage[dvipsnames]{xcolor}
\usepackage[utf8]{inputenc}
\usepackage{subfigure}
\usepackage{gensymb}
\usepackage{cleveref}
\usepackage{graphicx}
\usepackage{epstopdf}

\usepackage[T1]{fontenc}
\usepackage{pstricks}
\usepackage{color}
\usepackage{multirow}
\usepackage{slashed}
\usepackage{mathtools}

\usepackage[force]{feynmp-auto}
\usepackage{verbatim}

\newcommand{\Hub}{\mathcal{H}}

\newcommand{\s}{\mathfrak{s}}
\newcommand{\N}{\mathcal{N}}

\DeclareMathAlphabet{\mathpzc}{OT1}{pzc}{m}{it}

\newcommand{\SOTON}{Department of Physics and Astronomy, University of Southampton, SO17 1BJ Southampton, United Kingdom}

\begin{document}


\title{Leptogenesis in Type Ib seesaw models }


\author{Bowen Fu}
\email{B.Fu@soton.ac.uk}
\affiliation{\SOTON}

\author{Stephen F. King}
\email{king@soton.ac.uk}
\affiliation{\SOTON}

\date{\today}

\begin{abstract}
We study leptogenesis in three different realisations of the type Ib seesaw mechanism, where the effective masses of the neutrinos are obtained by the spontaneous symmetry breaking of two different Higgs doublets. In the minimal type Ib seesaw model, where two right-handed neutrinos form a Dirac mass, we show that it is impossible to produce the correct baryon asymmetry, even including a pseudo-Dirac mass splitting. In an extended type Ib seesaw model, with a third very heavy Majorana right-handed neutrino, together with the low scale Dirac pair of right-handed (RH) neutrinos and an extra singlet Higgs boson, we find that the imbalance of matter and antimatter can be explained by resonant leptogenesis. In the resulting low scale effective type Ib seesaw mechanism, we derive the allowed range of the seesaw couplings consistent with resonant leptogenesis. 
Dark matter may also be included via the right-handed neutrino portal.
The Dirac RH neutrino masses may lie in the 1-100 GeV mass range, accessible to the future experiments SHiP and FCC-$ee$, allowing the type Ib seesaw mechanism with leptogenesis and dark matter to be tested.
\end{abstract}


\maketitle

\section{Introduction}

Evidenced by the neutrino oscillation experiments \cite{2016NuPhB.908....1O}, the existence of neutrino masses and their mixing is a solid indication of new physics beyond the Standard Model (SM). In the last decades, hundreds of theories have been developed to explain the origin of the neutrino masses, most of which are various realisations of the dimension-five Weinberg operator \cite{Weinberg:1979sa}. Although the typical tree-level realisations of the Weinberg operator, including the type I \cite{Minkowski:1977sc,Yanagida:1979as,GellMann:1980vs,Mohapatra:1979ia}, II \cite{Magg:1980ut,Schechter:1980gr,Wetterich:1981bx,Lazarides:1980nt,Mohapatra:1980yp,Ma:1998dx} and III \cite{Foot:1988aq,Ma:1998dn,Ma:2002pf,Hambye:2003rt} seesaw models, are the most popular and well-studied among these theories, it is hard to generate proper neutrino mass naturally with large seesaw couplings and small right-handed (RH) neutrino masses simultaneously in these models, which makes experimental tests difficult. To make the models more testable, some low scale seesaw models with extended RH neutrino sectors such as the inverse seesaw model \cite{Mohapatra:1986bd} or the linear seesaw model \cite{Akhmedov:1995ip,Malinsky:2005bi} have been proposed. Alternatively, neutrino mass can also be generated radiatively through loop diagrams \cite{Zee:1980ai,Ma:2009dk,Bonnet:2012kz,Cai:2017jrq}.

As the seesaw mechanism involves lepton number violation, it is commonly linked to leptogenesis \cite{Fukugita:1986hr,Luty:1992un,Plumacher:1996kc,Covi:1996wh,Antusch:2004xy,King:2004wd,Buchmuller:2005eh,Antusch:2005tu,Davidson:2008bu,Bjorkeroth:2015tsa,King:2018fqh} which provides an attractive and minimal origin of the baryon asymmetry in the Universe. Using the observed value of Baryon asymmetry \cite{Aghanim:2018eyx}, the seesaw parameters can be constrained and related to other cosmological problems like the dark matter \cite{Gu:2007ug,Falkowski:2011xh,DiBari:2016guw,Chianese:2019epo,Liu:2020mxj,Chianese:2020khl,Chang:2021ose}. However, standard leptogenesis can hardly be related to collider experiments, as there is a well-known lower bound of RH neutrino mass around $10^9$ GeV \cite{Davidson:2002qv}. One way to reduce the lower bound is so-called resonant leptogenesis \cite{Pilaftsis:2003gt,Pilaftsis:2005rv,Xing:2006ms,Branco:2009by,Dev:2014laa,Ghiglieri:2018wbs,Granelli:2020ysj}, where the neutrino masses are quasi-degenerate with a mass splitting of order their decay rates. Moreover, the $B-L$ asymmetry can be enhanced by flavour effects since different processes come into equilibrium as the universe cools down and make the lepton flavours distinguishable \cite{Abada:2006fw,Nardi:2006fx,Blanchet:2006be,Antusch:2006cw,Dev:2017trv}. Nevertheless, the testability of type I seesaw leptogenesis remains an interesting topic in neutrino physics as well as cosmology \cite{Gu:2008yk,Fong:2013gaa,Drewes:2016jae,Dutta:2018zkg,Abada:2018oly}. 

\begin{figure}[t!]
\begin{center}
\includegraphics[height=0.15\textwidth]{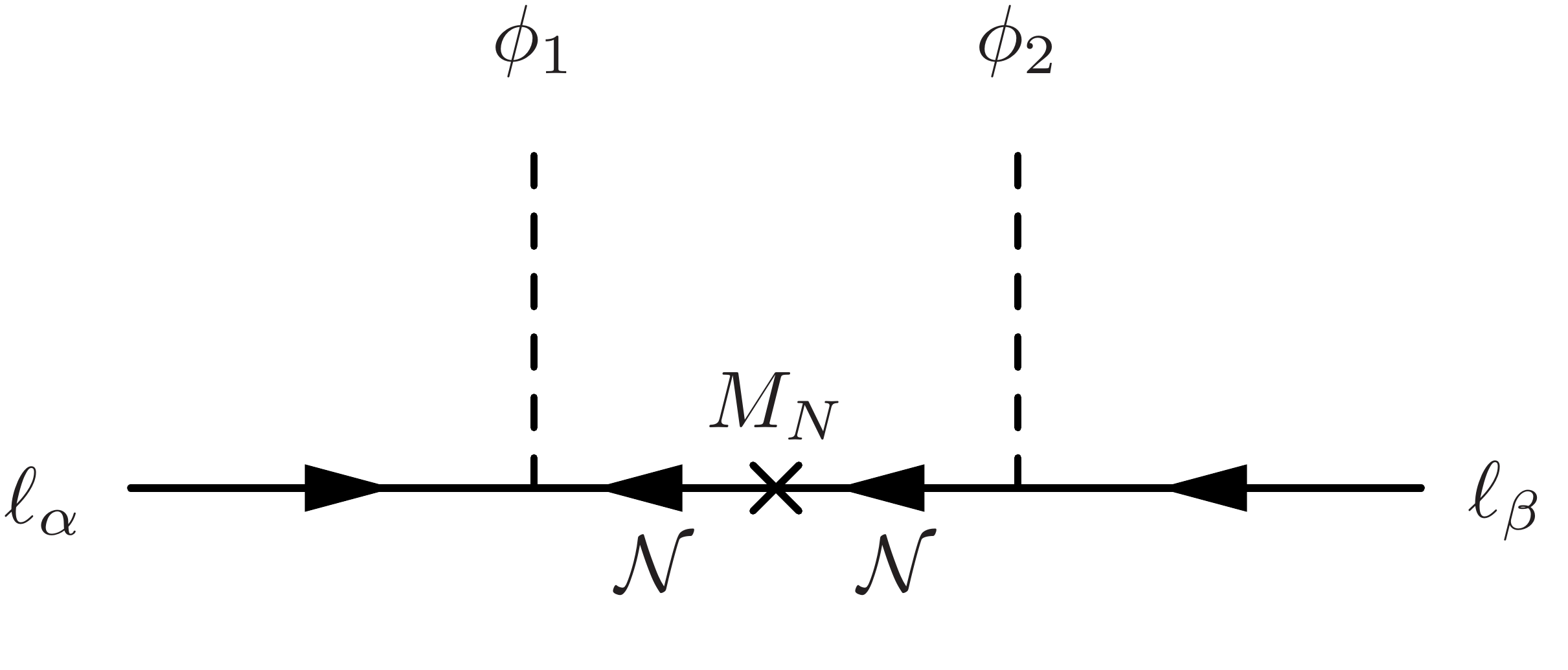}
\caption{\label{fig:NMIb} Neutrino mass in Type Ib seesaw model.}
\end{center}
\end{figure}

Recently a new version of the type I seesaw mechanism that can be just as testable as the low scale seesaw models above has been proposed, allowing just two RH neutrinos \cite{King:1999mb} in its minimal version, which is called the type Ib seesaw mechanism \cite{Hernandez-Garcia:2019uof}. Dark matter may also be included in the type Ib seesaw model via the right-handed neutrino portal \cite{Chianese:2021toe}. 
We shall sometimes refer to the usual type I seesaw mechanism as type Ia to distinguish it from the type Ib seesaw mechanism.
With multiple Higgs doublets, it had been noticed a long time ago that alternative pathways to the traditional Weinberg operator approach are possible in neutrino mass models \cite{Oliver:2001eg}. Systematic classification has been made to the two Higgs doublet models (2HDMs) by many authors \cite{Aoki:2009ha,Branco:2011iw,Chao:2012pt}, and the type Ib seesaw model is based on the so-called type II 2HDM where the down type quarks and charged leptons couple to one Higgs doublet and the up type quarks couples to the other one. Usually, only the Higgs doublets that couple to up type quarks is responsible for neutrino mass in the seesaw mechanism when the type I seesaw mechanism is combined with the type II 2HDM. However, as a novel feature of the type Ib seesaw mechanism, the effective neutrino mass operator involves both of the Higgs doublets coupling to up and down type quarks, while the two RH neutrinos form a single Dirac pair in the minimal case \cite{Chianese:2021toe}, as shown in Fig.\ref{fig:NMIb}. Different from the traditional type I seesaw models, the type Ib seesaw model allows a large seesaw coupling and a relatively small heavy neutrino mass simultaneously, making such models testable at colliders. The type Ib seesaw model shares many of the general features of testability as the inverse seesaw or linear seesaw models mentioned above, however it is distinguished by the simplicity of the RH neutrino sector, with 2 RHNs forming a single heavy Dirac mass in the minimal case, as mentioned above, rather than relying on extended RH neutrino sectors with more than 3 RHNs in other low energy seesaw models. 

In this paper, we discuss the leptogenesis in different realisations of the type Ib seesaw mechanism. In Sec.\ref{sec:Ib}, we start with the minimal type Ib seesaw model and find that the single low scale RH neutrino mass is incompatible with the current baryon asymmetry in such a model,
although a very heavy RH neutrino mass is possible. We then extend the minimal type Ib seesaw model with a small mass splitting in the RH neutrino mass in Sec.\ref{sec:Ib_pseu}. However, due to the special structure of Yukawa couplings in the model, resonant leptogenesis cannot be achieved even if flavour effects are taken into account. Finally, we turn to a new realisation of the model in Sec.\ref{sec:Ib_ex}, by including a third heavy right-handed neutrino \footnote{Introducing the third superheavy right-handed neutrino raises the puzzle of the RH neutrino mass hierarchy, which we shall not address here. In this paper we focus on low energy testability, noting that the third superheavy RH neutrino does not affect the testability of the low scale effective type Ib model.}, together with an extra singlet Higgs boson, leading to an effective and testable type Ib seesaw mechanism at low energy. In such a model, correct baryon asymmetry through resonant leptogenesis is shown to be available below TeV scale by simple approximated solutions, which constrains the larger seesaw Yukawa coupling, depending on the heavy neutrino mass. For RH neutrino mass below 100 GeV, the parameter space allowed by leptogenesis can be probed at the future experiments SHiP and FCC-$ee$, with dark matter also possible in the same region \cite{Chianese:2021toe}, allowing a simultaneous test of the type Ib seesaw mechanism, leptogenesis and dark matter.


\section{Minimal type Ib seesaw model \label{sec:Ib}}
As discussed in \cite{Chianese:2021toe}, the minimal version of the type Ib seesaw model involves two RH neutrinos $N_{R1},\,N_{R2}$ and two Higgs doublets $\phi_1,\phi_2$, where all the fields transform under a $Z_3$ symmetry in such as way as to require two different Higgs doublets in the seesaw mechanism. The charges of the particles under $SU(2)\times U(1)\times Z_3$ symmetry is shown in Tab.\ref{tab:sym_Ib}.\footnote{The $Z_3$ is allowed to be softly broken by terms in the scalar potential such as the Higgs mixing term $m_{12}\overline{\phi}_1\phi_2$ to avoid the domain wall problem \cite{Babu:2014pxa} and to satisfy the bound on the mass of the physical pseudoscalar \cite{ParticleDataGroup:2020ssz}.}
\begin{table}[t!]
\centering
\begin{tabular}{|c|c|c|c|c|c|c|c|c|c|}
\hline 
& ${q}_\alpha$ & ${u_R}_\beta$ & ${d_R}_\beta$
& $\ell_\alpha$ & ${e_R}_\beta$ & $\phi_{1}$ & $\phi_{2}$ & $N_{\mathrm{R}1}$ & $N_{\mathrm{R}2}$ \\[1pt] \hline 
$SU(2)_L$ & {\bf 2} & {\bf 1} & {\bf 1} & {\bf 2} & {\bf 1} & {\bf 2} & {\bf 2} & {\bf 1} & {\bf 1} \\ \hline & & & & & & & & & \\ [-1em]
$U(1)_Y$ & $\frac{1}{6}$ & $\frac{2}{3}$ & $-\frac{1}{3}$ & $-\frac{1}{2}$ & $-1$ & $-\frac{1}{2}$ & $-\frac{1}{2}$ & 0 & 0 \\[2pt] \hline & & & & & & & & & \\ [-1em]
$Z_3$ & $1$ & $\omega$ &  $\omega$ & $1$ & $\omega$ &  $\omega$ &  $\omega^2$ & $\omega^2$ &  $\omega$  \\ \hline 
\end{tabular}
\caption{\label{tab:sym_Ib}Irreducible representations of the fields of the model under the electroweak $SU(2)_L\times U(1)_Y$ gauge symmetry and the discrete $Z_3$ symmetry (where we write $\omega =e^{i2\pi/3}$). The fields $Q_{\alpha}, \ell_{\alpha}$ are left-handed SM doublets while ${u_R}_\beta,{d_R}_\beta,{e_R}_\beta$
are RH SM singlets where $\alpha, \beta = 1,2,3$ label the three families of quarks and leptons. The fields $N_{R1},N_{R2}$ are the two right-handed neutrinos.}
\end{table}
The $Z_3$ symmetry ensures that the coupling between the Higgs doublets and SM fermions follows the type II 2HDM pattern: The masses of the charged leptons and $-1/3$ charged quarks originate from the spontaneous symmetry breaking (SSB) of the first Higgs doublet $\phi_1$, while the $2/3$ charged quarks gain masses from $\phi_2$. The corresponding terms in the Lagrangian are
\begin{eqnarray}
\mathcal{L}_{\rm 2HDM} & \supset & - Y^u_{\alpha \beta} \overline{q}_\alpha { \phi}_2 u_{R\beta}
- Y^d_{\alpha \beta} \overline{q}_\alpha {\tilde \phi}_1 d_{R\beta} \nonumber\\&&- Y^e_{\alpha \beta} \overline{\ell}_\alpha {\tilde \phi}_1 e_{R\beta}
 + {\rm h.c.}\,
\label{eq:Yuk} 
\end{eqnarray}
In the type Ib seesaw sector, the Yukawa interactions take the form
\begin{eqnarray}
\mathcal{L}_{\rm seesawIb} & = & - Y_{1\alpha} \overline{\ell}_\alpha {\phi}_1 N_{R1} 
- Y_{2\alpha} \overline{\ell}_\alpha {\phi}_2 N_{R2}
\nonumber\\&& - M\overline{N^c_{R1}} N_{R2} + {\rm h.c.},
\label{eq:lg_Ib}
\end{eqnarray}
where $\ell_\alpha$ are the lepton doublets. The mass matrix of the RH neutrinos is constrained to be off-diagonal (i.e. Dirac),
\begin{eqnarray}
M_N &=&\begin{pmatrix} 0 & M \\[8pt] M & 0 \end{pmatrix} .
\label{MR}
\end{eqnarray}
The two ``right-handed'' Weyl neutrinos actually form a four-component Dirac spinor $\N = \left( N^c_{R1},N_{R2} \right)$ with a Dirac mass $M$. Alternatively, the neutrinos can be rotated into a Majorana basis through an orthogonal transformation \cite{Schechter:1980gr}
\begin{eqnarray}
\begin{pmatrix} N_{R1} \\[8pt] N_{R2} \end{pmatrix} \rightarrow 
\begin{pmatrix} n_{R1} \\[8pt] n_{R2} \end{pmatrix}=
\frac{1}{\sqrt{2}}\begin{pmatrix} 1& -1 \\[8pt] 1 & 1 \end{pmatrix} 
\begin{pmatrix} N_{R1} \\[8pt] N_{R2} \end{pmatrix}.
\end{eqnarray}
The diagonalised mass matrix is then $M'_N= \text{diag} (- M, M)$. After redefining the phase of $n_{R1}$, the Lagrangian of type Ib seesaw model in the Majorana basis $\left(n_{R1},n_{R2}\right)$ can be obtained by substituting $N_{R1}$ with $(-i\, n_{R1} + n_{R2})/\sqrt{2}$ and $N_{R2}$ with $(n_{R2} + i\, n_{R1})/\sqrt{2}$
\begin{eqnarray}
\mathcal{L}_{\rm seesawIb} & = & 
- \frac{i}{\sqrt{2}}\, \overline{\ell}_\alpha (Y_{2\alpha} {\phi}_2 - Y_{1\alpha} {\phi}_1) n_{R1} 
\nonumber\\ && - \frac{1}{\sqrt{2}}\, \overline{\ell}_\alpha (Y_{1\alpha} {\phi}_1 + Y_{2\alpha} {\phi}_2) n_{R2}
\nonumber\\ && - \frac12 M ( \overline{n^c_{R1}} n_{R1} + \overline{n^c_{R2}} n_{R2}) + {\rm h.c.}\label{eq:lg_Ib_M}
\\& = & 
- y_{ij\alpha}\overline{\ell}_\alpha n_{Ri} {\phi}_j 
- \frac12 M \, \overline{n^c_{Ri}} n_{Ri} + {\rm h.c.},
\label{eq:lg_Ib_M_2}
\end{eqnarray}
where the coupling $y_{ij\alpha}$ is defined as
\begin{eqnarray}
y_{1j\alpha} &=& \frac{i}{\sqrt{2}} (-1)^j Y_{j\alpha}, \quad y_{2j\alpha} = \frac{1}{\sqrt{2}} Y_{j\alpha}.\label{eq:newcouplings}
\end{eqnarray}
Following the procedure in \cite{Broncano:2002rw}, we integrate out the heavy neutrino to generate a set of effective operators, which leads to an effective field theory for the low energy phenomenology. The dimension-five effective operators are Weinberg-type operators involving 
the two different Higgs doublets $\phi_1,\phi_2$ \cite{Hernandez-Garcia:2019uof} are
\begin{eqnarray}
\delta \mathcal{L}^{d=5} &=& 
\frac{1}{2M}
\overline{\ell^c}_\alpha\left(Y_{1\alpha}^*Y_{2\beta}^*{\phi}_1^*{\phi}_2^\dagger +Y_{2\alpha}^*Y_{1\beta}^*{\phi}_2^*{\phi}_1^\dagger\right) \ell_\beta 
+ {\rm h.c.}\nonumber\\
\label{eq:d5op}
\end{eqnarray}
An alternative approach is to treat the Majorana mass term as a perturbative interaction where the two-component Weyl neutrinos form a four-component Dirac spinor $\N = \left( N^c_{R1},N_{R2} \right)$ with a Dirac mass $M$. The corresponding Feynman diagram is shown in Fig.\ref{fig:NMIb}. When the Higgs doublets develop vacuum expectation values (VEVs) as $\langle\phi_i\rangle=\left(v_i/\sqrt{2},0\right) $, the new Weinberg-type operator induces Majorana mass terms $m_{\alpha \beta} \nu_{\alpha} \nu_{\beta}$ for the light SM neutrinos, where 
\begin{eqnarray}
m_{\alpha \beta}= \dfrac{v_1 v_2}{2 M}\left(Y_{1\alpha}^*Y_{2\beta}^* + Y_{2\alpha}^*Y_{1\beta}^* \right).
\label{eq:dim5_relation_simp}
\end{eqnarray} 
In the case of a normal ordering (NO) which is favoured by the experimental results \cite{Capozzi:2017ipn}, the Yukawa couplings in the flavour basis are determined up to overall constants by the oscillation data as \cite{Hernandez-Garcia:2019uof}
\begin{eqnarray}
Y_{1\alpha} &=&\dfrac{Y_1}{\sqrt{2}}\left( \sqrt{1+\rho} \left(U_\text{PMNS}\right)_{\alpha 3}-\sqrt{1-\rho} \left(U_\text{PMNS}\right)_{\alpha 2}\right)\nonumber\\ &&\times e^{i\sigma}, 
\label{eq:Y1_NH}\\
Y_{2\alpha} &=& \dfrac{Y_2}{\sqrt{2}}\left( \sqrt{1+\rho} \left(U_\text{PMNS}\right)_{\alpha 3}+\sqrt{1-\rho} \left(U_\text{PMNS}\right)_{\alpha 2}\right)\nonumber\\ &&\times e^{-i\sigma}, 
\label{eq:Y2_NH}
\end{eqnarray}
where $\rho={(\sqrt{1+r}-\sqrt{r})}/{(\sqrt{1+r}+\sqrt{r})}$ with $r \equiv {\vert\Delta M_{21}^2\vert}/{\vert\Delta M_{32}^2\vert}$ and $Y_2,\,Y_1$ are real numbers. The phase $\sigma$ can be absorbed by the RH neutrino fields. After removing the phase, the couplings satisfy 
\begin{eqnarray}
\sum_{\alpha}Y_{i\alpha}^* Y_{i\alpha} &= & Y_i^2 \quad \text{and} \quad 
\sum_{\alpha}Y_{1\alpha}^* Y_{2\alpha} =  \rho Y_1Y_2.
\label{eq:relY}
\end{eqnarray} 
The neutrino masses in the normal hierarchy (NH) are
\begin{eqnarray}
m_1=0, &\ \ & \vert m_2 \vert=\frac{Y_1Y_2v_1v_2}{2M}(1-\rho),\\ \vert m_3\vert&=&\frac{Y_1Y_2v_1v_2}{2M}(1+\rho).
\end{eqnarray}
In 2HDMs, it is common to define the ratio of Higgs VEVs as $\tan \beta =v_2/v_1$, where $v_1$ and $v_2$ are the VEVs of $\phi_1$ and $\phi_2$ respectively. Assuming there is no complex relative phase between the VEVs, the Higgs VEVs follow the relation $\sqrt{v_1^2+v_2^2} = v = 246$ GeV and thus the VEVs can be expressed as $v_1=v\cos\beta$ and $v_2=v\sin\beta$. Using the central values of oscillation parameters \cite{Esteban:2020cvm}, a dimensionless combination is fixed as
\begin{eqnarray}
\frac{Y_1 Y_2 v\sin{\beta}\cos{\beta}}{M} = 2.4 \times 10^{-13}.
\label{eq:rel2}
\end{eqnarray}
In summary, there are four free parameters in the minimal type Ib seesaw model: two real seesaw Yukawa coupling parameters $Y_1,\,Y_2$, the heavy neutrino mass $M$ and $\tan\beta$, which are constrained by one relation Eq.\eqref{eq:rel2}. It is useful to derive a lower limit to the sum of squared Yukawa couplings from Eq.\eqref{eq:rel2} using the inequality of arithmetic and geometric means (AM–GM inequality)
\begin{eqnarray}
Y_1^2 + Y_2^2 \gtrsim 4.8 \times 10^{-13} \frac{M}{v\sin{\beta}\cos{\beta}}.
\label{eq:rel3}
\end{eqnarray}

\subsection{CP asymmetry \label{sec:Ib_asym}}

\begin{figure}[t!]
\begin{center}
\subfigure[]{ \includegraphics[height=0.15\textwidth]{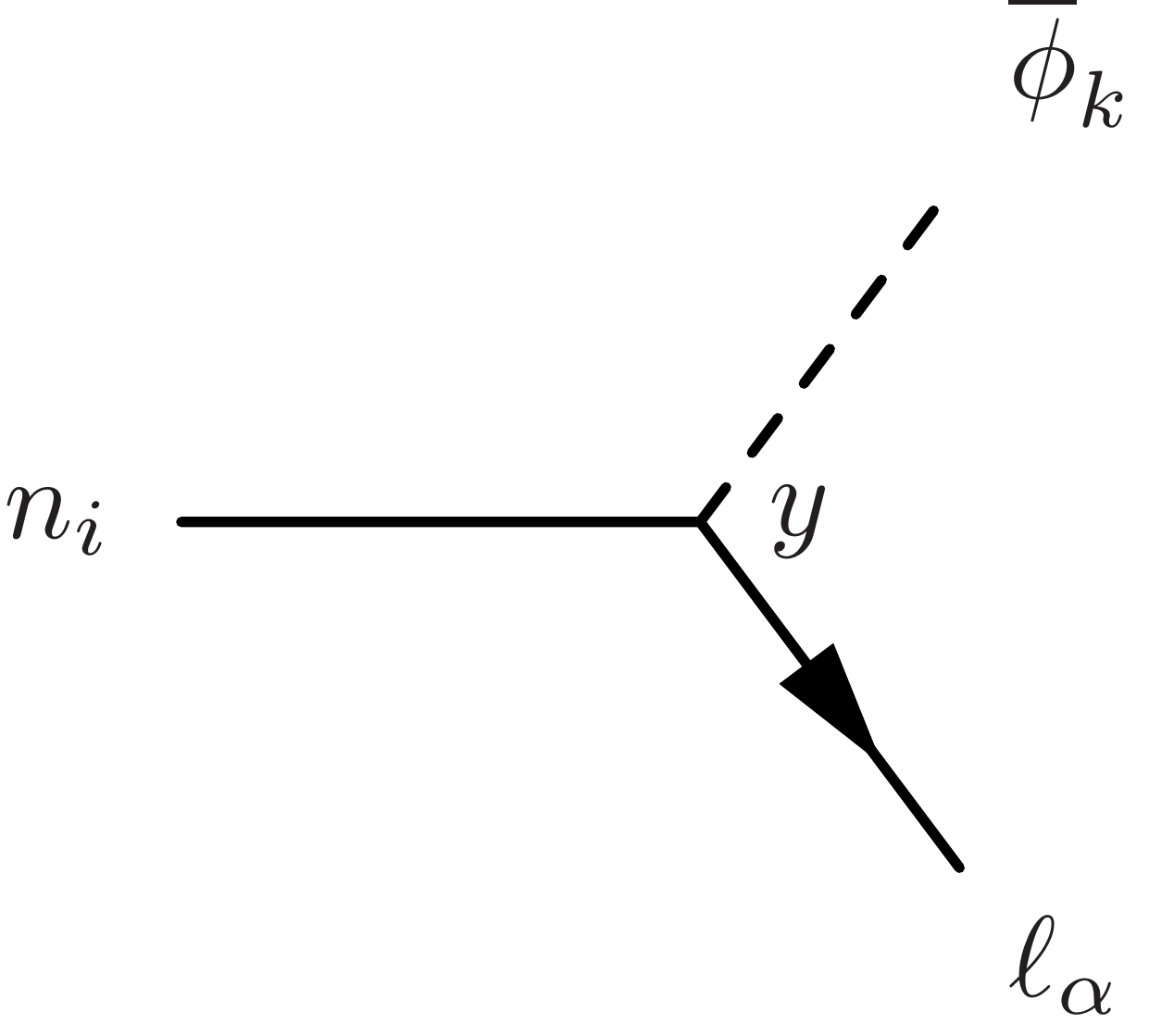} \label{fig:tree}}
\subfigure[]{ \includegraphics[height=0.15\textwidth]{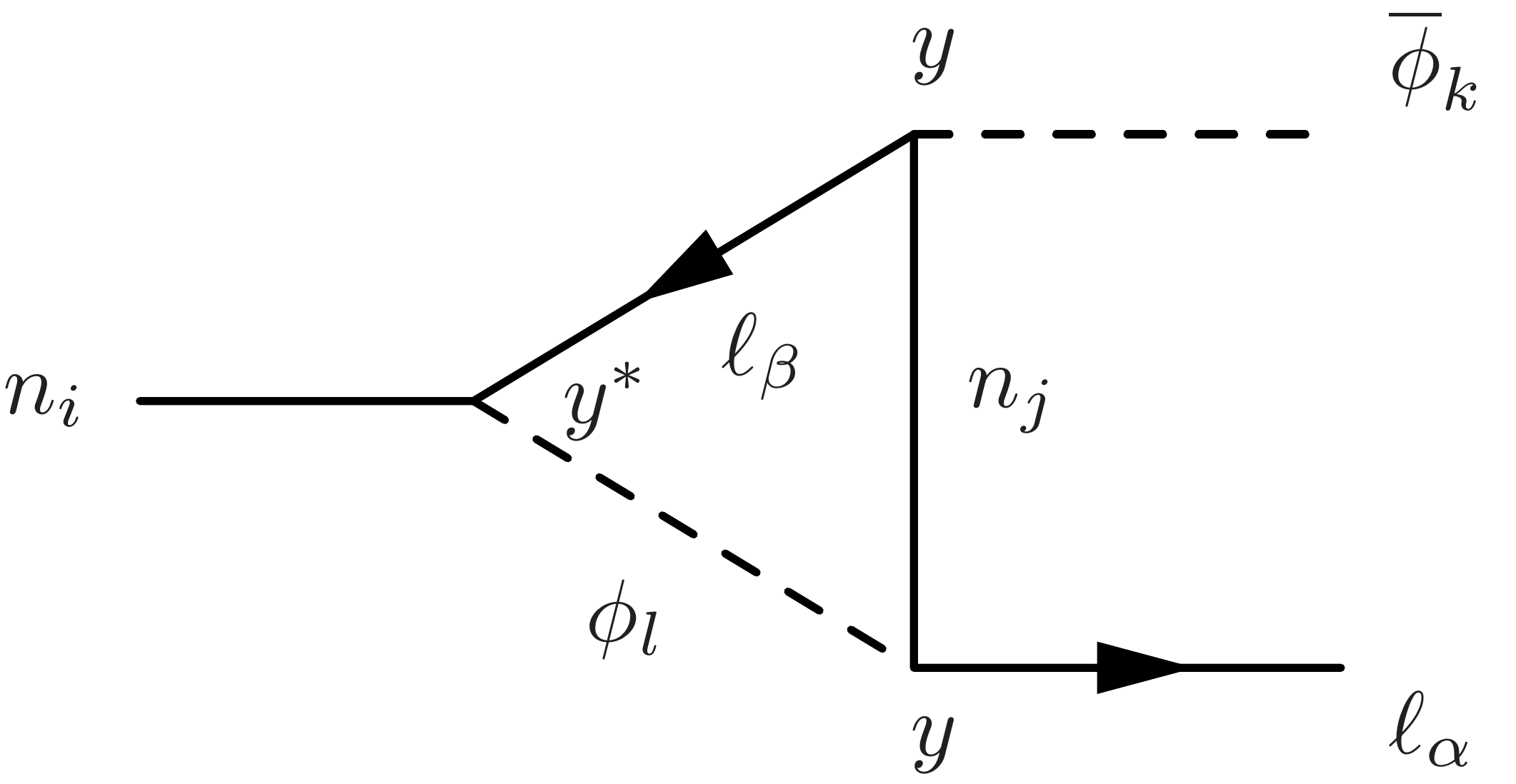} \label{fig:vertex}}
\subfigure[]{ \includegraphics[height=0.15\textwidth]{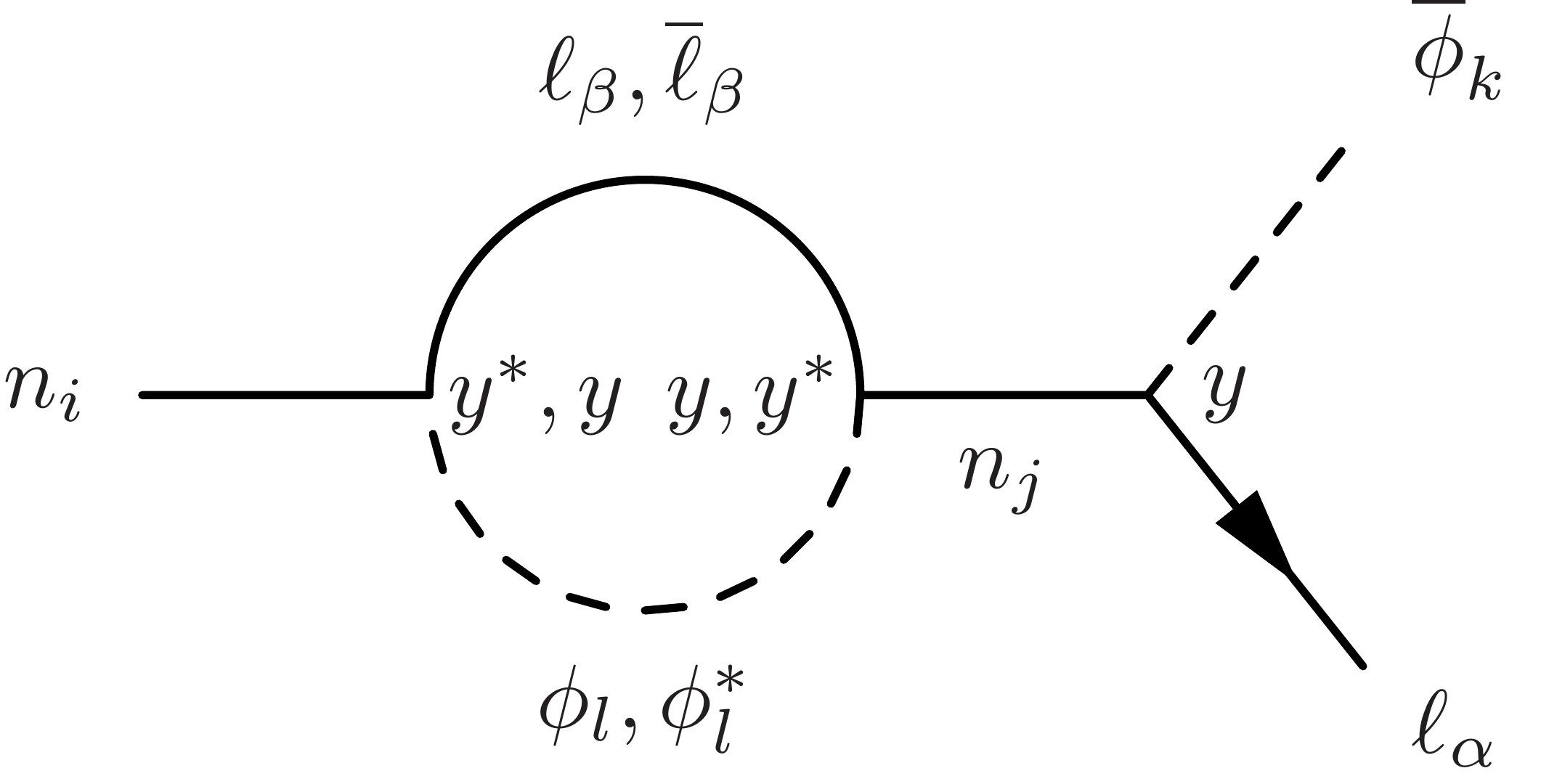} \label{fig:wave}}
\caption{tree-level and 1-loop level decay of the heavy neutrinos}
\end{center}
\end{figure}

In the type Ib seesaw model, the RH neutrinos can decay into different Higgs doublets and the tree-level decay rates for the two RH neutrinos are  
\begin{eqnarray}
\Gamma^{\rm tree}_{n_1\ell}&=&\Gamma^{\rm tree}_{n_1\overline{\ell}}=\Gamma^{\rm tree}_{n_2\ell}= \Gamma^{\rm tree}_{n_2\overline{\ell}}=\frac{Y_1^2 + Y_2^2}{32\pi}M.
\end{eqnarray} 
The quantity characterising CP asymmetry is defined as
\begin{eqnarray}
\epsilon_{n_i} =\frac{\Gamma_{n_i\ell} - \Gamma_{n_i\overline{\ell}}}{\Gamma_{n_i\ell} + \Gamma_{n_i\overline{\ell}}} 
= \epsilon_{n_i}^{\rm vertex} + \epsilon_{n_i}^{\rm wave},
\end{eqnarray} 
where $\epsilon_{n_i}^{\rm vertex}$ ($\epsilon_{n_i}^{\rm wave}$) stands for the contribution from the interference between the tree-level diagram in Fig.\ref{fig:tree} and the vertex (wave-function) diagram in Fig.\ref{fig:vertex} (Fig.\ref{fig:wave}). As the two heavy neutrinos have a degenerate mass, the wave-function contribution has to vanish. The contributions from vertex diagram is given by \cite{Covi:1996wh,Antusch:2006cw}
\begin{eqnarray}
\left(\epsilon_{n_1}^{\rm vertex}\right)_{k\alpha}&\equiv&
\frac{\Gamma_{n_1\rightarrow \ell_\alpha\phi_k}^{\rm vertex} - \Gamma_{n_1\rightarrow\overline{\ell}_\alpha\phi_k^\dagger}^{\rm vertex}}{\Gamma_{n_1\ell} + \Gamma_{n_1\overline{\ell}}}
\nonumber\\ &=& -\frac{\Im\left(\sum_{j,l,\beta} y_{1l\beta}^*y_{jk\beta}y_{jl\alpha}y_{1k\alpha}^*\right)}{8\pi\sum_{k,\alpha} \left(y_{1k\alpha}y_{1k\alpha}^*\right)} f\left(1\right),
\label{eq:epsvn1}\\
\left(\epsilon_{n_2}^{\rm vertex}\right)_{k\alpha}&\equiv&
\frac{\Gamma_{n_2\rightarrow \ell_\alpha\phi_k}^{\rm vertex} - \Gamma_{n_2\rightarrow\overline{\ell}_\alpha\phi_k^\dagger}^{\rm vertex}}{\Gamma_{n_2\ell} + \Gamma_{n_2\overline{\ell}}} 
\nonumber\\ &=& -\frac{\Im\left(\sum_{j,l,\beta} y_{2l\beta}^*y_{jk\beta}y_{jl\alpha}y_{2k\alpha}^*\right)}{8\pi\sum_{k,\alpha} \left(y_{2k\alpha}y_{2k\alpha}^*\right)} f\left(1\right),\quad
\label{eq:epsvn2}
\end{eqnarray} 
where $f(x)=\sqrt{x}(1-(1+x)\ln[(1+x)/x])$. The asymmetry here is not only flavour ($\alpha$) dependent but also Higgs ($k$) dependent because the two Higgs doublets are distinguishable as only $\phi_2$ couples to the top quark \footnote{The Higgs doublets are assumed to be in their electroweak eigenstates. As will be shown later, leptogenesis can only be achieved at a very high temperature in this scenario, where the mass mixing of the Higgs bosons is neglectable compared with the thermal effects.}. Using Eq.\eqref{eq:newcouplings} and Eq.\eqref{eq:relY}, 
the above numerator summations can be expressed as
\begin{eqnarray}
\sum_{j,l,\beta} y_{1l\beta}^*y_{jk\beta}y_{jl\alpha}y_{1k\alpha}^* 
&= & \sum_{l,\beta} \frac{1+(-1)^{k+l+1}}{4}Y_{l\beta}^*Y_{k\beta}Y_{l\alpha}Y_{k\alpha}^*
\nonumber\\ &=& \sum_{l} \frac{\rho Y_1Y_2}{2}\varepsilon_{lk} Y_{l\alpha}Y_{k\alpha}^*,\\
\sum_{j,l,\beta} y_{2l\beta}^*y_{jk\beta}y_{jl\alpha}y_{2k\alpha}^* 
&= & \sum_{l,\beta} \frac{1+(-1)^{k+l+1}}{4}Y_{l\beta}^*Y_{k\beta}Y_{l\alpha}Y_{k\alpha}^*
\nonumber\\ &=& \sum_{l} \frac{\rho Y_1Y_2}{2}\varepsilon_{lk} Y_{l\alpha}Y_{k\alpha}^*,
\end{eqnarray} 
where $\varepsilon_{11}=\varepsilon_{22}=0$ and $\varepsilon_{12}=\varepsilon_{21}=1$. So the numerators of Eq.\eqref{eq:epsvn1} and Eq.\eqref{eq:epsvn2} are 
\begin{eqnarray}
&&\Im\left(\sum_{l} \frac{\rho Y_1Y_2}{2}\varepsilon_{lk} Y_{l\alpha}Y_{k\alpha}^* \right)\nonumber\\&&\quad\quad\quad\quad\quad\quad=
\begin{dcases}
\frac{\rho Y_1Y_2}{2}\Im\left(Y_{2\alpha}Y_{1\alpha}^*\right), & k=1\,;\\
\frac{\rho Y_1Y_2}{2}\Im\left(Y_{1\alpha}Y_{2\alpha}^*\right), & k=2.
\end{dcases}\quad\quad\quad
\end{eqnarray} 
Notice that the contributions to the asymmetry from diagrams with different Higgs doublets in the final states have opposite signs. If the Higgs doublets are indistinguishable, the Higgs index $k$ must be summed over and the total asymmetry vanishes. The numerical value of the numerators of the fractions in Eq.\eqref{eq:epsvn1} and Eq.\eqref{eq:epsvn2} reads 
\begin{eqnarray}
&&\Im\left(\sum_{j,l,\beta} y_{1l\beta}^*y_{jk\beta}y_{jl\alpha}y_{1k\alpha}^* \right)=
\Im\left(\sum_{j,l,\beta} y_{2l\beta}^*y_{jk\beta}y_{jl\alpha}y_{2k\alpha}^* \right)\nonumber\\&&=
10^{-2}\sin\delta\, Y_1^2Y_2^2
\begin{pmatrix} 
-2.03& 1.16 & 0.87 \\
 2.03&-1.16 &-0.87
\end{pmatrix}_{k\alpha},
\end{eqnarray} 
where $\delta$ is the Dirac phase in the PMNS matrix. Notice that the relative Majorana phase between the second and third neutrinos can also affect the quantity above. However, we assume the Majorana phase is 0 for the simplicity of this work, since introducing such a phase does not change the result significantly in order of magnitude. If the Higgs doublets or all of the neutrino flavours are indistinguishable, the Higgs index or the flavour index has to be summed over and the asymmetry contributed by the vertex process vanishes. Only when both the Higgs doublets and the neutrino flavours are distinguishable, the asymmetry survives. The asymmetries in lepton flavour $\alpha$ associated with Higgs doublet $\phi_k$ are
\begin{eqnarray}
&&\left(\epsilon_{n_1}^{\rm vertex}\right)_{k\alpha}=\left(\epsilon_{n_2}^{\rm vertex}\right)_{k\alpha}\nonumber \\&&=
-10^{-4}\sin\delta\, \frac{Y_1^2Y_2^2}{Y_1^2+Y_2^2} 
\begin{pmatrix} 
6.24& -3.56 & -2.69 \\
-6.24& 3.56 & 2.69
\end{pmatrix}_{k\alpha}.
\end{eqnarray} 
The CP violation in scatterings was first taken into account in the early works on resonant leptogenesis \cite{Pilaftsis:2003gt,Pilaftsis:2005rv}, assuming factorisation, see also
\cite{Abada:2006ea}. When the washout is particularly weak and the lepton asymmetries are produced at high temperature, the CP violation from gauge scatterings becomes relevant \cite{Nardi:2007jp,Fong:2010bh}. However, it is shown later that the washout is very strong in the type Ib seesaw model, and so the factorisation treatment is a viable approximation in this scenario.

\subsection{Boltzmann equations and the approximate solution \label{sec:Ib_BE}}

In the type Ib seesaw model, both of the two RH neutrinos contribute to leptogenesis. The Boltzmann equation for the RH neutrinos is
\begin{eqnarray}
\frac{dY_{n_i}}{dz} = - \frac{z}{\s\Hub(M)}&& \left(\frac{Y_{n_i}}{Y^{\rm eq}_{n_i}} - 1\right)
\nonumber \\&&\times\left( \gamma_{n_i\rightarrow2} + \gamma_{2\rightarrow2}^{n_iA} + \gamma_{2\rightarrow2}^{n_it} \right),
\label{eq:BE_ni_sim}
\end{eqnarray} 
where $z=M/T$. $\gamma_{n_i\rightarrow2}$, $\gamma_{2\rightarrow2}^{n_iA}$ and $\gamma_{2\rightarrow2}^{n_it}$ are the thermal averaged two-body decay rate, $2\rightarrow2$ scattering rate with gauge bosons and $2\rightarrow2$ scattering rate with top quark. The Boltzmann equation for the $B-L_\alpha$ asymmetry can be separated into a source term and a washout term
\begin{eqnarray}
\frac{dY_{\Delta_{\alpha}}}{dz} &=& \left(\frac{dY_{\Delta_{\alpha}}}{dz}\right)^s + \left(\frac{dY_{\Delta_{\alpha}}}{dz}\right)^w , \label{eq:BE:parts}
\end{eqnarray} 
where 
\begin{eqnarray}
&&\left(\frac{dY_{\Delta_{\alpha}}}{dz}\right)^s\simeq -\frac{z}{s\Hub(M)} \sum_{i}\left(\frac{Y_{n_i}}{Y^{\rm eq}_{n_i}} - 1\right)\gamma_{2\rightarrow2}^{n_it} \left(\epsilon_{n_i}\right)_{2\alpha}
\nonumber\\&& = \sum_{i} \frac{dY_{n_i}}{dz} \left(\epsilon_{n_i}\right)_{2\alpha} \frac{\gamma_{2\rightarrow2}^{n_it}}{\gamma_{n_i\rightarrow2} + \gamma_{2\rightarrow2}^{n_iA} + \gamma_{2\rightarrow2}^{n_it}}.
\end{eqnarray}
Unlike in the general case where the two-body decay and $2\rightarrow2$ scattering processes with gauge bosons also contribute to the source of $B-L_\alpha$ asymmetry, there is only the contribution from the $2\rightarrow2$ scattering processes with top quark in the type Ib seesaw model with degenerate RH neutrinos as the two Higgs doublets are not distinguishable in the other processes. In the $2\rightarrow2$ scattering processes with the top quark, only $\phi_2$ can serve as a mediator because the top quark does not couple to $\phi_1$. Further discussion about the Boltzmann equations can be found in Appendix \ref{ap:BE}. Here we would like to adopt the approximation made in \cite{Abada:2006ea}, where the contribution to washout from the scattering processes is taken into account by mutiplying a temperature dependent factor.

As discussed in many papers \cite{Barbieri:1999ma,Abada:2006ea,Antusch:2006cw,Davidson:2008bu,King:2018fqh} the approximate solution of the Boltzmann equation for $B/3-L_\alpha$ asymmetry can be estimated by the product of CP asymmetry in neutrino decay $\epsilon_\alpha$ and the efficiency factor $\eta_\alpha$. A generalisation of such solution with contribution from different heavy neutrinos is
\begin{eqnarray}
Y_{\Delta\alpha} =\sum_i \epsilon_{i\alpha} \eta_{i\alpha} Y^{\rm eq}_{n_i}(z\ll1),
\end{eqnarray}
where the approximate expression of $\eta_{i\alpha}$ depends on the strength of washout. To discuss how strong the washout is, it is useful to define dimensional quantities $\tilde{m}_i$ and $m^*_i$ as 
\begin{eqnarray} 
\tilde{m}_i &=& 8\pi\frac{v_i^2}{M^2}\Gamma_{n_i} = \frac{v_i^2}{2M}\left(Y_1^2 + Y_2^2\right),\\
m^*_i &=& 4\pi\frac{v_i^2}{M^2}\Hub(M)\simeq 1.1\times10^{-3}\, \text{eV}\, \frac{v_i^2}{v^2},\label{eq:m_quantities}
\end{eqnarray} 
where $v = 246$ GeV is the VEV of the SM Higgs. A flavour dependent $\tilde{m}_{i\alpha}$ can be defined as 
\begin{eqnarray} 
\tilde{m}_{i\alpha} = v_i^2(|Y_{1\alpha}|^2 + |Y_{2\alpha}|^2)/2M.\label{eq:m_quantities_flavor}
\end{eqnarray} 
The ``strong washout'' scenario appears when the rate of inverse decay is fast compared to the Hubble parameter when the RH neutrinos start to decay, i.e. $\Gamma_{n_i}>\Hub(M)$, which is equivalent to $\tilde{m}_i > m^*_i$ and $\left(Y_1^2 + Y_2^2\right)> 8.9 \times10^{-15}M/v$. Notice that such condition has to be satisfied in type Ib seesaw model as $\left(Y_1^2 + Y_2^2\right)$ is constrained by the neutrino data through Eq.\eqref{eq:rel3}. In such a scenario, the neutrino $n_i$ has a thermal number density at temperature $T\sim M$, and any asymmetry produced by the decays is washed out until the inverse decays become out of equilibrium. For lepton flavour $\alpha$, the asymmetry surviving depends on the decay rate of RH neutrinos into $\ell_\alpha$. To estimate the efficiency factor $\eta_{i\alpha}$ presenting the effects from ``washout'' processes and inefficiency in RH neutrino production, it is required to compare $m^*_i$ with flavour dependent $\tilde{m}_{i\alpha}$ for flavour $\alpha$. For the different lepton flavours, $\tilde{m}_{i\alpha}$ are proportional to 
\begin{eqnarray} 
|Y_{1e}|^2 + |Y_{2e}|^2 &=& 0.063 \left(Y_1^2 + Y_2^2\right) \nonumber\\&-& 0.057\, \cos\delta \left(Y_1^2-Y_2^2\right)\nonumber\\&\gtrsim& 0.005\left(Y_1^2 + Y_2^2\right),\\
|Y_{1\mu}|^2 + |Y_{2\mu}|^2 &=& (0.520-0.010\cos\delta) \left(Y_1^2 + Y_2^2\right) \nonumber\\&+& (-0.289+0.033\cos\delta) \left(Y_1^2-Y_2^2\right)\nonumber\\&\gtrsim& 0.208\left(Y_1^2 + Y_2^2\right),\\
|Y_{1\tau}|^2 + |Y_{2\tau}|^2 &=& (0.417+0.010\cos\delta) \left(Y_1^2 + Y_2^2\right) \nonumber\\&+& (0.289+0.025\cos\delta) \left(Y_1^2-Y_2^2\right)\nonumber\\&\gtrsim& 0.114\left(Y_1^2 + Y_2^2\right).
\end{eqnarray}
Together with Eq.\eqref{eq:rel3}, the upper limits of the ratio between $m^*_i$ and $\tilde{m}_{i\alpha}$ are
\begin{eqnarray} 
\frac{m^*_i}{\tilde{m}_{ie}} \lesssim 3.7\,&& \sin{\beta}\cos{\beta} ,\quad
\frac{m^*_i}{\tilde{m}_{i\mu}} \lesssim 0.09\, \sin{\beta}\cos{\beta} ,\nonumber\\&&
\frac{m^*_i}{\tilde{m}_{i\tau}} \lesssim 0.13\, \sin{\beta}\cos{\beta} .
\end{eqnarray} 
For $\mu$ and $\tau$ leptons, there must be $\tilde{m}_{i\alpha} > m^*_i$ and the asymmetry in these lepton flavours survive after all the washout processes become out of equilibrium. The corresponding efficiency factor is roughly \cite{Abada:2006ea}
\begin{eqnarray} 
\eta_{i\alpha} \simeq \left(\frac{m^*_i}{2\left|A_{\alpha\alpha}\right|\tilde{m}_{i\alpha}}\right)^{1.16} \simeq \frac{m^*_i}{2\left|A_{\alpha\alpha}\right|\tilde{m}_{i\alpha}},
\label{eq:eta}
\end{eqnarray} 
where $A$ is the so-called ``$A$-matrix'' parameterising the effect of interactions in thermal equilibrium \cite{Antusch:2006cw}. Detailed calculations about $A$ in type Ib seesaw model can be found in Appendix \ref{ap:Amatrix}. For $e$ leptons, $\tilde{m}_{ie}$ can be smaller than $m^*_i$ if $\tan\beta<3.7$ \footnote{In the case $\tan\beta < 3.7$, the washout process is strong overall, but it is weak in the electron flavour. In that case, the RH neutrinos can be brought into thermal equilibrium by the inverse decay as in the case of strong washout but the asymmetry produced at $T>M$ in electron flavour is not completely washed out. As a result, the efficiency factor in the electron flavour is different from that in the strong washout scenario. For further discussion, see \cite{Abada:2006ea,Davidson:2008bu}.}, otherwise the $\tilde{m}_{i\alpha}$ has to be larger than $m^*_i$ due to the neutrino data. If $\tan\beta>3.7$, then $\tilde{m}_{ie} > m^*_i$ and the corresponding efficiency factor share the same expression as $\mu$ and $\tau$ flavours. The $B-L_\alpha$ asymmetry can be estimated by
\begin{eqnarray}
Y_{\Delta\alpha} &=& \sum_i \left(\epsilon_{n_i}\right)_{2\alpha} \eta_{i\alpha} \frac{\gamma_{2\rightarrow2}^{n_it}}{\gamma_{n_i\rightarrow2} + \gamma_{2\rightarrow2}^{n_iA} + \gamma_{2\rightarrow2}^{n_it}} Y^{\rm eq}_{n_i}(z\ll1) \nonumber\\
&\simeq& -4 \times 10^{-21}\sin\delta
\begin{pmatrix} 
6.24& -3.56 & -2.69 \\
-6.24& 3.56 & 2.69
\end{pmatrix}_{2\alpha} 
\nonumber\\&&\times\left|A_{\alpha\alpha}\right|^{-1}\frac{M}{v} \left(|Y_{1\alpha}|^2 + |Y_{2\alpha}|^2\right)^{-1}\nonumber\\&&\times
\frac{Y_1^2Y_2^2}{Y_1^2+Y_2^2}\sum_i \frac{\gamma_{2\rightarrow2}^{n_it}}{\gamma_{n_i\rightarrow2} + \gamma_{2\rightarrow2}^{n_iA} + \gamma_{2\rightarrow2}^{n_it}}.
\end{eqnarray}
With the relation Eq.\eqref{eq:rel2}, the limit for the asymmetry in each flavour can be derived as 
\begin{eqnarray}
\left|Y_{\Delta e}\right| &\lesssim& 9 \times 10^{-20} \left|A_{ee}\right|^{-1} \frac{M}{v} ,
\\
\left|Y_{\Delta \mu}\right| &\lesssim& -7 \times 10^{-21} \left|A_{\mu\mu}\right|^{-1} \frac{M}{v} ,
\\
\left|Y_{\Delta \tau}\right| &\lesssim& -7 \times 10^{-21} \left|A_{\tau\tau}\right|^{-1} \frac{M}{v} .
\end{eqnarray}
The result only requires $Y_1Y_2$ to be a constant but its value is irrelevant. The sum of the asymmetries is too small for $M<(1+\tan\beta^2)\times10^9$ GeV to satisfy the current observed value of baryogenesis \cite{Aghanim:2018eyx}
\begin{eqnarray}
Y_{\Delta B} &\sim& \frac{10}{31}\sum_\alpha Y_{\Delta\alpha} \simeq 8.7 \times10^{-11}.\label{eq:baryo}
\end{eqnarray}
Therefore we assume that only $\ell_\tau$ is distinguishable. In that case, the $B-L_\alpha$ asymmetry is given by
\begin{eqnarray}
\left|Y_{\Delta e+\mu}\right| &\lesssim& 6 \times 10^{-20} \frac{M}{v} ,\\
\left|Y_{\Delta \tau}\right| &\lesssim& - 10^{-20} \frac{M}{v} ,
\end{eqnarray}
and Eq.\eqref{eq:baryo} gives $M > 10^{12}$ GeV. In this case, the lower limit of the RH neutrino mass is close to its upper limit for flavour dependent leptogenesis, i.e. allowed range of the RH neutrino mass is very constrained. Although such a scenario is possible,
such a superheavy RH neutrino is unlikely to have any testable effects, which goes against the original motivation for the type Ib seesaw model. Therefore we are motivated to consider a modification of the model in order to realise correct leptogenesis and light RH neutrinos simultaneously.


\section{Type Ib seesaw model with a pseudo-Dirac neutrino \label{sec:Ib_pseu}}

\begin{figure}[t!]
\begin{center}
\subfigure[]{ \includegraphics[height=0.15\textwidth]{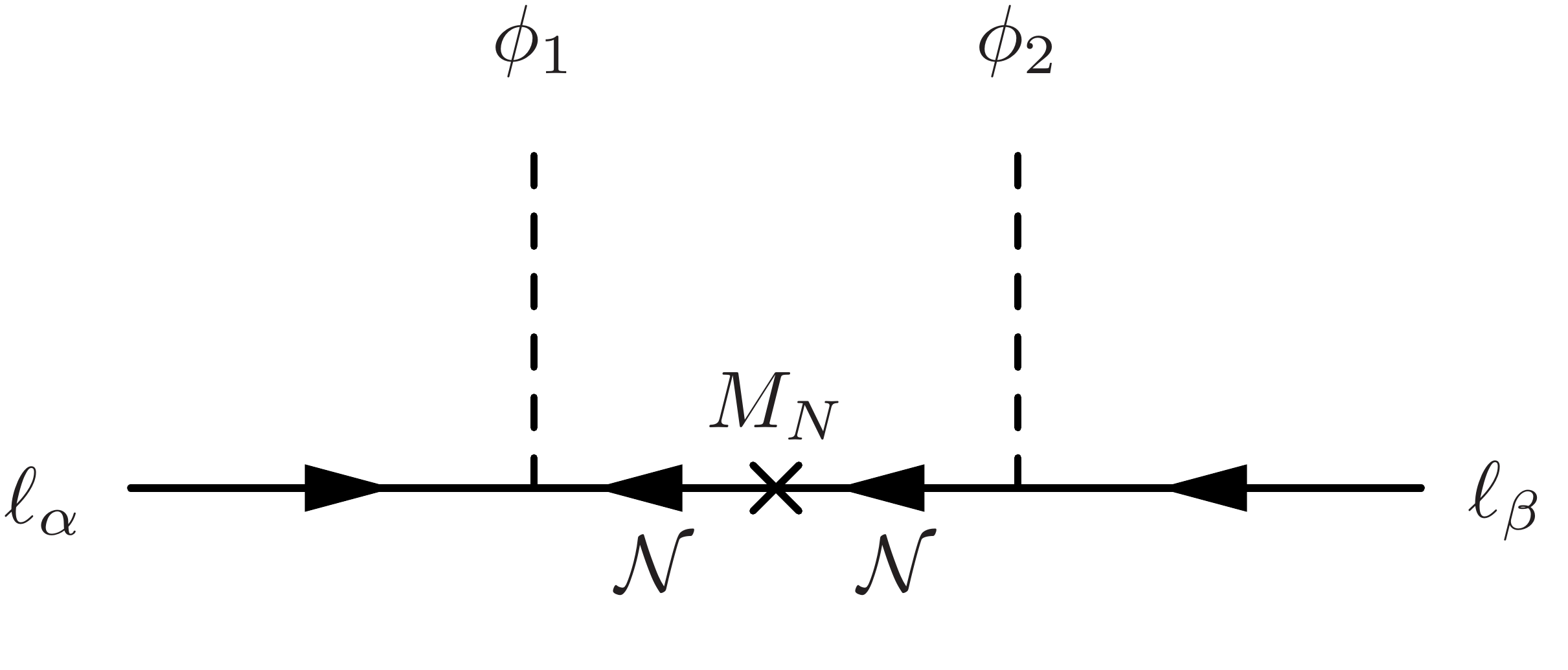} \label{fig:NMdirac}}
\subfigure[]{ \includegraphics[height=0.15\textwidth]{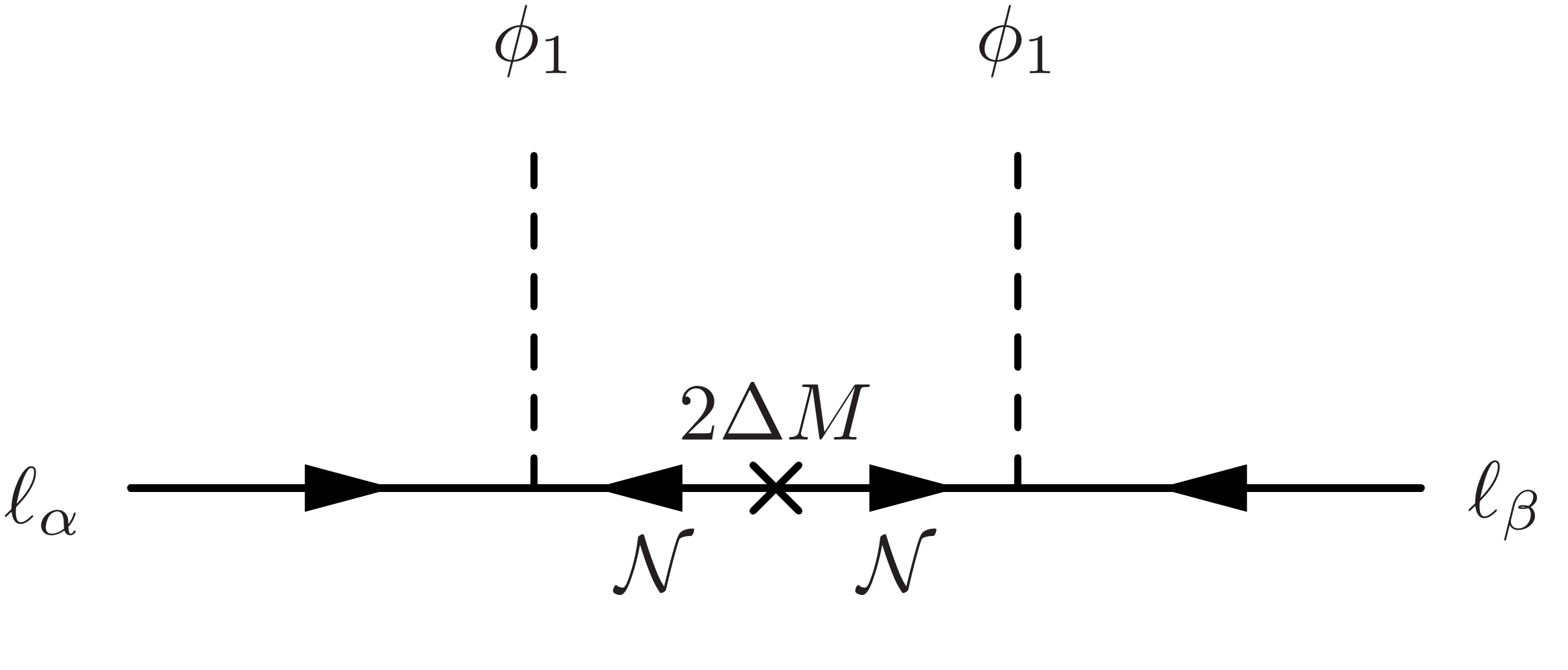} \label{fig:NMmajor}}
\caption{ Light neutrino mass generated by (a) the type Ib seesaw mechanism involving two different Higgs doublets $\phi_1$ and $\phi_2$ and (b) the traditional type Ia seesaw mechanism involving the same Higgs doublet $\phi_1$.}
\end{center}
\end{figure}

In order to possibly realise leptogenesis at a lower energy scale in the type Ib seesaw model, in this section
we try introducing a Majorana mass for one of the RH neutrinos so that the neutrinos become quasi-Dirac. In principle, both of the neutrino Majorana masses can softly break the $Z_3$ symmetry. For simplicity, the Majorana mass of $N_{R1}$ is assumed to be zero while the one of $N_{R2}$ is $\Delta M$. The smallness of the Majorana mass is ensured by the softness of $Z_3$ symmetry breaking. Then the Lagrangian becomes
\begin{eqnarray}
\mathcal{L}_{\rm seesawIb} & = & - Y_{1\alpha} \overline{\ell}_\alpha {\phi}_1 N_{R1} 
- Y_{2\alpha} \overline{\ell}_\alpha {\phi}_2 N_{R2}
\nonumber\\&&- M\overline{N^c_{R1}} N_{R2} - \Delta M\,\overline{N^c_{R2}} N_{R2} + {\rm h.c.}.\quad\quad
\label{eq:lg_sd}
\end{eqnarray}
Similar treatment can be applied to change the RH neutrinos from their Dirac basis to Majorana basis $\left(n_{R1},n_{R2}\right)$ and the Lagrangian becomes
\begin{eqnarray}
\mathcal{L}_{\rm seesawIb} & = & 
- i\, \overline{\ell}_\alpha (Y_{2\alpha} {\phi}_2\sin\theta - Y_{1\alpha} {\phi}_1\cos\theta) n_{R1} \nonumber\\&&
- \overline{\ell}_\alpha (Y_{1\alpha} {\phi}_1\sin\theta + Y_{2\alpha} {\phi}_2\cos\theta) n_{R2}\nonumber\\&&
- \frac12 (M - \Delta M)\, \overline{n^c_{R1}} n_{R1} \nonumber\\&&- \frac12 (M + \Delta M) \,\overline{n^c_{R2}} n_{R2} + {\rm h.c.}
\\& \equiv & 
- y_{ij\alpha}\overline{\ell}_\alpha n_{Ri} {\phi}_j 
- \frac12 M_i \, \overline{n^c_{Ri}} n_{Ri} + {\rm h.c.},\quad\quad
\label{eq:couplings_sd}
\end{eqnarray}
where the couplings $y_{ij\alpha}$ and masses $M_i$ are properly defined by the above equivalences. Different from the exact degenerate neutrino case, where the rotation angle is exactly $\pi/4$, the rotation angle here satisfies $\sin2\theta\sim1$ and $\cos2\theta\sim\Delta M/M$ when $\Delta M/M\ll1$, which leads to $\sin\theta\sim(1- \Delta M/2M)/\sqrt2$ and $\cos\theta\sim(1+ \Delta M/2M)/\sqrt2$. The dimension-five effective operators in this case are \cite{Hernandez-Garcia:2019uof}
\begin{eqnarray}
\delta \mathcal{L}^{d=5} &=& 
\frac{1}{2M_1}y_{1j\alpha}^*y_{1k\beta}^*\overline{\ell^c}_\alpha {\phi}_j^* {\phi}_k^\dagger \ell_\beta \nonumber\\&&
+\frac{1}{2M_2}y_{2j\alpha}^*y_{2k\beta}^*\overline{\ell^c}_\alpha {\phi}_j^* {\phi}_k^\dagger \ell_\beta 
+ {\rm h.c.}
\\&\simeq& 
\frac12\frac{1}{M}
\overline{\ell^c}_\alpha \left[ - Y_{1\alpha}^* Y_{1\beta}^* {\phi}_1^* {\phi}_1^\dagger \frac{2\Delta M}{M} \right. \nonumber\\&&\left.
+ \left(Y_{1\alpha}^*Y_{2\beta}^*{\phi}_1^*{\phi}_2^\dagger +Y_{2\alpha}^*Y_{1\beta}^*{\phi}_2^*{\phi}_1^\dagger\right)\right] \ell_\beta 
+ {\rm h.c.}\quad\quad
\label{eq:d5op_sd}
\end{eqnarray}
where we apply the approximation $\sin2\theta\sim1$ and $\cos2\theta\sim\Delta M/M$ when $\Delta M/M\ll1$. The same result can be worked out in the Dirac basis and the light neutrino mass is generated by both Fig.\ref{fig:NMdirac} and Fig.\ref{fig:NMmajor}. 
The light neutrino mass generated in Fig.\ref{fig:NMdirac} is proportional to $1/M$ and the light neutrino mass generated in Fig.\ref{fig:NMmajor} is proportional to $\Delta M/M^2$ \footnote{Notice that although the mass splitting appears as an extra mass term for $N_{R2}$ coupling to $\phi_2$ in the Lagrangian, the Dirac propagator ``brings'' a $N_{R2}$ to $N_{R1}$ in the seesaw limit. Therefore the Higgs doublet evolved in the dimension-five effective operator is actually $\phi_1$ instead of $\phi_2$.}. At leading order, the effective operators in Eq.\eqref{eq:d5op_sd} agree with the ones in Eq.\eqref{eq:d5op} for the type Ib seesaw model with exactly degenerate RH neutrino mass. 

\subsection{CP asymmetry for resonant leptogenesis \label{sec:Ib_pseu_asym}}
 As we aim to focus on GeV scale heavy neutrinos, the thermal effect is important before the EW sphaleron decouples. At high temperture, the themal mass of the Higgs boson can be large so that the Higgs boson decays into heavy neutrino and charged leptons. Instead of the heavy neutrino decay, the Higgs decay can produce the asymmetry in leptons which can be converted into baryogenesis.

With a mass splitting in the RH neutrino mass, the decay rate of the RH neutrinos are 
\begin{eqnarray}
\Gamma^{\rm tree}_{n_1\ell}&=&\Gamma^{\rm tree}_{n_1\overline{\ell}}=\frac{Y_1^2\cos^2\theta + Y_2^2\sin^2\theta}{16\pi}(M-\Delta M) , \quad\\
\Gamma^{\rm tree}_{n_2\ell}&=&\Gamma^{\rm tree}_{n_2\overline{\ell}}=\frac{Y_1^2\sin^2\theta + Y_2^2\cos^2\theta}{16\pi}(M+\Delta M) ,\quad
\end{eqnarray} 
both of which have the same leading order limit $Y_1^2M/16\pi$. In the resonant leptogenesis regime, the CP asymmetry is dominantly produced from the decay of heavy neutrinos with self-energy. The corresponding contributions are given by 
\begin{eqnarray}
\left(\epsilon_{n_1}^{\rm wave}\right)_{k\alpha}&\equiv&
\frac{\Gamma_{n_1\rightarrow \ell_\alpha\phi_k}^{\rm wave} - \Gamma_{n_1\rightarrow\overline{\ell}_\alpha\phi_k^\dagger}^{\rm wave}}{\Gamma_{n_1\ell} + \Gamma_{n_1\overline{\ell}}} \nonumber\\
&=& \frac{ - \Im\left[\left(y_{1k\alpha}y_{2k\alpha}^*\right) \sum_{l,\beta} \left(y_{1l\beta}y_{2l\beta}^*\right)\right]}{\sum_{k,\alpha} \left(y_{1k\alpha}y_{1k\alpha}^*\right) \sum_{l,\beta} \left(y_{2l\beta}y_{2l\beta}^*\right)} 
\nonumber\\&&\times\frac{2 \Delta M_{21} \Gamma_{n_2}}{4\Delta M^2_{21} + \Gamma_{n_2}^2},
\label{eq:epswn1}\\
\left(\epsilon_{n_2}^{\rm wave}\right)_{k\alpha}&\equiv&
\frac{\Gamma_{n_2\rightarrow \ell_\alpha\phi_k}^{\rm wave} - \Gamma_{n_2\rightarrow\overline{\ell}_\alpha\phi_k^\dagger}^{\rm wave}}{\Gamma_{n_2\ell} + \Gamma_{n_2\overline{\ell}}} \nonumber\\
&=&\frac{- \Im\left[\left(y_{1k\alpha}y_{2k\alpha}^*\right) \sum_{l,\beta} \left(y_{1l\beta}y_{2l\beta}^*\right) \right]}{\sum_{k,\alpha} \left(y_{1k\alpha}y_{1k\alpha}^*\right) \sum_{l,\beta} \left(y_{2l\beta}y_{2l\beta}^*\right)} 
\nonumber\\&&\times\frac{2 \Delta M_{21} \Gamma_{n_1}}{4\Delta M^2_{21} + \Gamma_{n_1}^2},
\label{eq:epswn2}
\end{eqnarray} 
where $\Delta M_{21} = M_2 - M_1$. The last fraction in Eq.\eqref{eq:epswn1} (or Eq.\eqref{eq:epswn2}) is small unless $\Delta M_{21} \sim \Gamma_{n_i}/2$ and it equals $1/2$ when $\Delta M_{21} = \Gamma_{n_i}/2$. The relation $\Delta M_{21} = \Gamma_{n_i}/2$ is generally refered to as the resonance condition. Notice that the couplings defined in Eq.\eqref{eq:couplings_sd} satisfy
\begin{eqnarray}
y_{1j\beta}y_{2j\beta}^* & = & i (-1)^jY_{j\beta}Y_{j\beta}^*\cos\theta\sin\theta .
\end{eqnarray}
This quantity, whether sum over the indices or not, is completely imaginary. As a result, the numerators of the first fraction in Eq.\eqref{eq:epswn1} and Eq.\eqref{eq:epswn2} have to vanish. This is a property of the type Ib seesaw model due to its special structure. In general, the RH neutrinos can have Yukawa couplings to both of the Higgs doublets. In the type Ia seesaw model with a mass splitting, as discussed in \cite{Zhao:2021tgi}, there is a contribution to resonant leptogenesis which is proportional to $Y_{1\beta}Y_{2\beta}^*$ and it can survive. However, in the type Ib seesaw model, the coupling to each Higgs doublet is unique. As $y_{1j\beta}$ and $y_{2j\beta}^*$ have the same Higgs index, their product is proportional to $Y_{j\beta}Y_{j\beta}^*$ which is naturally real \footnote{The phase $\sigma$ in Eq.\eqref{eq:Y1_NH} and Eq.\eqref{eq:Y2_NH} can also contribute to the resonant leptogenesis in type Ia as discussed in \cite{Zhao:2021tgi}. However, this also requires the product of couplings with different Higgs index. When the couplings with the same Higgs index appear as a pair $y y^*$, the phases are always cancelled.}. Therefore the asymmetry factors in Eq.\eqref{eq:epswn1} and Eq.\eqref{eq:epswn2} are zero. Since the contribution to lepton asymmetry from both the decay and scattering process are proportional to the same asymmetry factor, we conclude that the minimal type Ib seesaw model with a pseudo-Dirac heavy neutrino is not consistent with resonant leptogenesis.

This result can be interpreted in a more general way. To have a non-zero contribution from the wave-function diagram, not only is the non-degenerate RH neutrino mass required, but the Yukawa couplings also have some constraints on their phase. Suppose the neutrinos are coupling to a group of Higgs fields $\phi_k$ through general Yukawa terms $e^{i\theta_{ik\alpha}}Y_{ik\alpha}\overline{\ell_\alpha} \phi_k n_{R_i}$ where $Y_{\alpha i}\in \mathbb{R}$. The flavour dependent contribution from the wave-function process is proportional to 
\begin{widetext}
\begin{eqnarray}
\left(\epsilon_{n_i}^{\rm wave-function}\right)_{k\alpha}&\propto&\sum_{j\neq i}\Im\left[ \left(e^{i\theta_{ik\alpha}}Y_{ik\alpha} e^{-i\theta_{jk\alpha}}Y_{jk\alpha}\right)
\sum_{l,\beta} \left(e^{i\theta_{il\beta}}Y_{il\beta} e^{-i\theta_{jl\beta}}Y_{jl\beta}\right)\right]\\
&\propto&\sum_{j\neq i}\sum_{l,\beta}Y_{ik\alpha}Y_{jk\alpha}Y_{il\beta} Y_{jl\beta} \sin\left(\theta_{ik\alpha}-\theta_{jk\alpha}+\theta_{il\beta}-\theta_{jl\beta}\right). \quad\quad\label{eq:asym_general}
\end{eqnarray} 
\end{widetext}
The contribution from the wave-function process vanishes if $\left(\theta_{ik\alpha}-\theta_{jk\alpha}+\theta_{il\beta}-\theta_{jl\beta}\right) \in \mathbb{Z}\pi$.

\section{Extended type Ib seesaw mechanism and its low energy effective theory\label{sec:Ib_ex}}

To find a possible realisation of the low scale type Ib seesaw mechanism which is compatible with resonant leptogenesis, we discuss a model with a third RH neutrino $N_{R3}$ with a large Majorana mass $M_{33}\gg M$
and a Higgs singlet $\xi$ where the type Ib seesaw Lagrangian can be obtained effectively. The symmetries of particles in such model are shown in Tab.\ref{tab:Ib_ex}.
\begin{table}[t!]
\centering
\begin{tabular}{|c|c|c|c|c|c|c|c|c|c|c|c|}
\hline 
& ${q}_\alpha$ & ${u_R}_\beta$ & ${d_R}_\beta$ & $\ell_\alpha$ & ${e_R}_\beta$ & $\phi_{1}$ & $\phi_{2}$ & $N_{\mathrm{R}1}$ & $N_{\mathrm{R}2}$ & $N_{\mathrm{R}3}$ & $\xi$\\[1pt] \hline & & & & & & & & & & & \\ [-1em]
$SU(2)_L$ & {\bf 2} & {\bf 1} & {\bf 1} & {\bf 2} & {\bf 1} & {\bf 2} & {\bf 2} & {\bf 1} & {\bf 1} & {\bf 1} & {\bf 1} \\ \hline & & & & & & & & & & & \\ [-1em]& & & & & & & & & & & \\ [-1em]
$U(1)_Y$ & $\frac{1}{6}$ & $\frac{2}{3}$ & $-\frac{1}{3}$ & $-\frac{1}{2}$ & $-1$ & $-\frac{1}{2}$ & $-\frac{1}{2}$ & 0 & 0 & 0 & 0 \\[2pt] \hline & & & & & & & & & & & \\ [-1em]
$Z_4$ & $1$ & $\omega^2$ &  $\omega$ & $1$ & $\omega$ &  $\omega$ &  $\omega^2$ & $\omega^3$ &  $\omega$ & $\omega^2$ & $\omega$ \\ \hline  
\end{tabular}
\caption{\label{tab:Ib_ex}Irreducible representations of the fields of the model under the electroweak $SU(2)_L\times U(1)_Y$ gauge symmetry and the discrete symmetries (where $\omega =e^{i\pi/2}$). The fields $Q_{\alpha}, L_{\alpha}$ are left-handed SM doublets while ${u_R}_\beta,{d_R}_\beta,{e_R}_\beta$
are right-handed SM singlets where $\alpha, \beta$ label the three families of quarks and leptons. The fields $N_{\mathrm{R}{1,2}}$ are the two right-handed neutrinos.}
\end{table}
The particles are charged under a $Z_4$ symmetry which can be broken by the VEV of the new Higgs singlet $\xi$. The interaction between Higgs doublets and charged fermions keeps the same structure as in the type II 2HDM 
\begin{eqnarray}
\mathcal{L}_{\rm 2HDM} \supset - Y^u_{\alpha \beta} \overline{q}_\alpha { \phi}_2 u_{R\beta}
&-& Y^d_{\alpha \beta} \overline{q}_\alpha {\tilde \phi}_1 d_{R\beta} \nonumber\\&-& Y^e_{\alpha \beta} \overline{\ell}_\alpha {\tilde \phi}_1 e_{R\beta}
 + {\rm h.c.}.\quad
\label{eq:Yuk_ex} 
\end{eqnarray}
while the seesaw Lagrangian reads
\begin{eqnarray}
\mathcal{L}_{\rm seesaw} & = & - Y_{1\alpha} \overline{\ell}_\alpha {\phi}_1 N_{R1} - Y_{3\alpha} \overline{\ell}_\alpha {\phi}_2 N_{R3} \nonumber\\&& - 2 Y_{13} \overline{\xi}\, \overline{N^c_{R3}} N_{R1} - 2 Y_{23} \xi \overline{N^c_{R3}} N_{R2} \nonumber\\&& 
- M\overline{N^c_{R1}} N_{R2} - \frac12 M_{33}\overline{N^c_{R3}} N_{R3} + {\rm h.c.},
\label{eq:lg_Ibex}
\end{eqnarray}
where the factor $2$ in the Yukawa interactions between RH neutrinos is introduced as a convenient convention. The mass matrix of the RH neutrinos, before the Higgs singlet VEV, is 
\begin{eqnarray}
M_N &=&\begin{pmatrix} 0 & M & 0 \\[8pt] M & 0 & 0 \\[8pt] 0 & 0 & M_{33} \end{pmatrix} .
\label{eq:mass_matrix}
\end{eqnarray}
At low energy, the singlet Higgs can gain a VEV $v_\xi$ and breaks the $Z_4$ symmetry. If $M_{33}\gg M$, two dimension-five effective operators can be generated after the third RH neutrino is integrated out \footnote{In general, there can be other dimension-five operators proportional to $\Lambda^{-1}$, where $\Lambda$ is some mass scale higher than $M_{33}$. The couplings of such dimension-five operators are more suppressed than $M_{33}^{-1}$ and therefore can be neglected. }
\begin{eqnarray}
- \frac{2Y_{13}Y_{3\alpha}}{M_{33}}\overline{\xi}\overline{\ell}_\alpha {\phi}_2 N_{R1} - \frac{2Y_{23}Y_{3\alpha}}{M_{33}}\xi\overline{\ell}_\alpha {\phi}_2 N_{R2} + {\rm h.c.}.\quad
\label{eq:eff_dim5}
\end{eqnarray}
As the $Z_4$ symmetry is broken by the VEV of singlet Higgs $\xi$ \footnote{The symmetry breaking would lead to a Goldstone boson from $\xi$. However, the coupling between such a Goldstone boson and SM particles is very weak as the scalar field only takes part in interactions involving $N_{R3}$. After $N_{R3}$ decouples, the effective interactions between the Goldstone boson and other particles are always suppressed by the mass of $N_{R3}$ as shown in Eq.\eqref{eq:eff_dim5}. Therefore we do not expect the Goldstone boson to cause any significant phenomenological effects. Of course any possible phenomenological problems could be alleviated by adding soft symmetry breaking terms to the scalar potential.}, the effective Lagrangian of the Yukawa interactions between neutrinos and Higgs doublets, below the scale of the Higgs singlet VEV, can be summarised as 
\begin{eqnarray}
\mathcal{L}_{\rm Yukawa}^{\rm eff} = - Y_{1\alpha} \overline{\ell}_\alpha {\phi}_1 N_{R1} &-& Y'_{1\alpha}\overline{\ell}_\alpha {\phi}_2 N_{R1} \nonumber\\&-& Y'_{2\alpha} \overline{\ell}_\alpha {\phi}_2 N_{R2} + {\rm h.c.}, \label{eq:Yuk_ef}\quad
\end{eqnarray}
where $Y'_{1\alpha}=2Y_{13}Y_{3\alpha} v_\xi/M_{33}$ and $Y'_{2\alpha}= 2Y_{23}Y_{3\alpha} v_\xi/M_{33}$. As $Y'_{1\alpha}$ and $Y'_{2\alpha}$ are suppressed by $v_\xi/M_{33}$, they are treated as relatively small when compared with the Yukawa couplings in Eq.\eqref{eq:lg_Ibex}. At leading order, the dimension-five operator from the type Ib seesaw mechanism is
\begin{eqnarray}
-\frac{1}{M}Y_{1\alpha}^*(Y'_{2\beta})^* \overline{\ell^c}_\alpha{\phi}_1^*{\phi}_2^\dagger \ell_\beta + {\rm h.c.}.
\end{eqnarray}
Notice that one of the Yukawa couplings is naturally small in this model, allowing a low scale Dirac mass $M$, possibly in the 1-100 GeV mass range accessible to experiments.

For the convenience of computing lepton asymmetries, we would like to work in the RH neutrino mass eigenstate basis, so the above Lagrangian must be rewritten in this basis. 
After the scalar singlet 
$\xi$ gains a VEV, the Lagrangian becomes
\begin{eqnarray}
\mathcal{L}_{\rm seesaw} & = & - Y_{1\alpha} \overline{\ell}_\alpha {\phi}_1 N_{R1} - Y_{3\alpha} \overline{\ell}_\alpha {\phi}_2 N_{R3} \nonumber\\&&- Y_{13} v_\xi\, \overline{N^c_{R3}} N_{R1} - Y_{23} v_\xi \overline{N^c_{R3}} N_{R2} \nonumber\\&&
- M\overline{N^c_{R1}} N_{R2} - \frac12 M_{33}\overline{N^c_{R3}} N_{R3} + {\rm h.c.}
\end{eqnarray}
The RH neutrino mass matrix becomes 
\begin{eqnarray}
M_N &=&\begin{pmatrix} 0 & M & M_{13} \\[8pt] M & 0 & M_{23} \\[8pt] M_{13} & M_{23} & M_{33} \end{pmatrix},
\label{eq:mass_matrix}
\end{eqnarray}
where $M_{13} = Y_{13}v_\xi$ and $M_{23} = Y_{23}v_\xi$. Such mass matrix can be diagonalised by unitary transformation. In the limit $M_{33} \gg M,\,|M_{13}|,\,|M_{23}|$, the transformation takes the form
\begin{widetext}
\begin{eqnarray}
\begin{pmatrix} N_{R1} \\[8pt] N_{R2} \\[8pt] N_{R3} \end{pmatrix} \rightarrow 
\begin{pmatrix} n_{R1} \\[8pt] n_{R2} \\[8pt] n_{R3} \end{pmatrix}=
\begin{pmatrix} 
\frac{1}{\sqrt{2}}\left(1- \frac{M_{13}^2-M_{23}^2}{4 M M_{33}}\right)^* & 
\frac{1}{\sqrt{2}}\left(1+ \frac{M_{13}^2-M_{23}^2}{4 M M_{33}}\right)^* & 
-\frac{M_{13}^*+M_{23}^*}{M_{33}}\\[8pt] 
\frac{i}{\sqrt{2}}\left(1+ \frac{M_{13}^2-M_{23}^2}{4 M M_{33}}\right)^* &
\frac{-i}{\sqrt{2}}\left(1- \frac{M_{13}^2-M_{23}^2}{4 M M_{33}}\right)^* &
-i\frac{M_{13}^*-M_{23}^*}{M_{33}} \\[8pt] 
\frac{M_{13}}{\sqrt{2}M_{33}} & \frac{M_{23}}{\sqrt{2}M_{33}} & 1 
\end{pmatrix} 
\begin{pmatrix} N_{R1} \\[8pt] N_{R2} \\[8pt] N_{R3} \end{pmatrix},
\end{eqnarray}
\end{widetext}
and the neutrino mass eigenvalues are
\begin{eqnarray}
M_1 &\simeq& M - \frac{\left(M_{13} + M_{23}\right)^2}{2M_{33}},\nonumber\\
M_2 \simeq M &+& \frac{\left(M_{13} - M_{23}\right)^2}{2M_{33}},\quad
M_3 \simeq M_{33} .
\label{eq:masses}
\end{eqnarray}
In general, $M_{13}$ and $M_{23}$ can be complex \footnote{The complexity of $M_{13}$ and $M_{23}$ cannot be removed by redefining the phase of $N_{Ri}$. In total, there are three phases in $N_{Ri}$. Two of them are fixed by making $M$ and $M_{33}$ in Eq.\eqref{eq:lg_Ibex} real and the other one is fixed by absorbing the seesaw coupling phase $\sigma$ in Eq.\eqref{eq:Y1_NH} and Eq.\eqref{eq:Y2_NH}.} as well as $M_1$ and $M_2$, and the complexity can be removed by redefining the RH neutrino fields $n_{R1}$ and $n_{R2}$ with $e^{i\theta_1/2}n_{R1}$ and $e^{i\theta_2/2}n_{R2}$. In fact, as will be shown later, at least one of the phases is required to be nonzero to obtain resonant leptogenesis, otherwise the contribution from the wave-function diagram would be zero due to the structure of the couplings. To realise a scenario with resonant leptogenesis, the mass splitting between $n_{R1}$ and $n_{R1}$ should be small, i.e. $|M_1|\simeq|M_2|\gg||M_1|-|M_2||$. To satisfy such condition while keeping $N_{R1}$ and $N_{R2}$ a pseudo-Dirac pair, it is assumed that $|M_{13}^2|,|M_{23}^2|\ll M M_{33}$ \footnote{Such condition can be satisfied in several ways. For example, $|M_{13}^2|\gg|M_{23}^2|,M M_{33}$ also make the mass splitting small, with a leading order mass $|M_{13}^2|$. However, in that case the pseudo-Dirac structure of $N_{R1}$ and $N_{R2}$ no longer exists.}.
Then the phases can be expressed as 
\begin{eqnarray}
\tan\theta_1&=&\frac{\Im \left(M_1\right)}{\Re \left(M_1\right)}\simeq - \frac{1}{2MM_{33}}\Im \left[\left(M_{13} + M_{23}\right)^2\right],\quad\\
\tan\theta_2&=&\frac{\Im \left(M_2\right)}{\Re \left(M_2\right)}\simeq \frac{1}{2MM_{33}}\Im \left[\left(M_{13} - M_{23}\right)^2\right].
\end{eqnarray}
and the transformation becomes
\begin{widetext}
\begin{eqnarray}
\begin{pmatrix} N_{R1} \\[8pt] N_{R2} \\[8pt] N_{R3} \end{pmatrix} \rightarrow 
\begin{pmatrix} n_{R1} \\[8pt] n_{R2} \\[8pt] n_{R3} \end{pmatrix}= 
\begin{pmatrix} 
\frac{1}{\sqrt{2}} e^{i\theta_1/2} & \frac{1}{\sqrt{2}} e^{i\theta_1/2} & -\frac{M_{13}^*+M_{23}^*}{M_{33}}e^{i\theta_1/2} \\[8pt] 
\frac{i}{\sqrt{2}} e^{i\theta_2/2} & -\frac{i}{\sqrt{2}} e^{i\theta_2/2} & -i\frac{M_{13}^*-M_{23}^*}{M_{33}} e^{i\theta_2/2} \\[8pt] 
\frac{M_{13}}{\sqrt{2}M_{33}} & \frac{M_{23}}{\sqrt{2}M_{33}} & 1 
\end{pmatrix} 
\begin{pmatrix} N_{R1} \\[8pt] N_{R2} \\[8pt] N_{R3} \end{pmatrix}.
\end{eqnarray}
\end{widetext}
After the transformation, the neutrino mass eigenvalues are
\begin{eqnarray}
M_1 &\simeq& M - \frac{\Re\left[\left(M_{13} + M_{23}\right)^2\right]}{2M_{33}},\nonumber\\
M_2 \simeq M &+& \frac{\Re\left[\left(M_{13} - M_{23}\right)^2\right]}{2M_{33}},\quad
M_3 \simeq M_{33} .
\label{eq:masses_re}
\end{eqnarray}
The Lagrangian of the extended type Ib seesaw model in the neutrino mass eigenstates $\left(n_{R1},\,n_{R2},\,n_{R3}\right)$ can be obtained 
\begin{eqnarray}
\mathcal{L}_{\rm seesaw} & = & 
- \frac12 M_1 \overline{n^c_{R1}} n_{R1} - \frac12 M_2 \overline{n^c_{R2}} n_{R2} - \frac12 M_3 \overline{n^c_{R3}} n_{R3}\nonumber\\&&
- \frac{1}{\sqrt{2}} \overline{\ell}_\alpha \left[Y_{1\alpha}\phi_1 - \left(Y'_{1\alpha}+Y'_{2\alpha}\right) \phi_2 \right] e^{-i\theta_1/2} n_{R1} 
\nonumber\\&&
- \frac{i}{\sqrt{2}}\overline{\ell}_\alpha \left[ - Y_{1\alpha} {\phi}_1 + \left(Y'_{1\alpha}-Y'_{2\alpha}\right) \phi_2\right]e^{-i\theta_2/2} n_{R2} 
\nonumber\\&&- \overline{\ell}_\alpha \left(Y_{13}^*Y_{1\alpha} \frac{v_\xi}{M_{33}} \phi_1 + Y_{3\alpha}\phi_2\right)n_{R3}
+ {\rm h.c.}
\label{eq:lg_Ib_2}
\\& \equiv & 
- \frac12 M_i \, \overline{n^c_{Ri}} n_{Ri} - y_{ij\alpha}\overline{\ell}_\alpha \phi_j n_{Ri} + {\rm h.c.},
\label{eq:lg_Ib_3}
\end{eqnarray}
where the coupling $y_{ij\alpha}$ reads
\begin{eqnarray}
y_{ij\alpha} = 
\begin{pmatrix} 
\frac{1}{\sqrt{2}} Y_{1\alpha} e^{-i\theta_1/2} & -\frac{1}{\sqrt{2}} \left(Y'_{1\alpha}+Y'_{2\alpha}\right) e^{-i\theta_1/2} \\[8pt] 
-\frac{i}{\sqrt{2}} Y_{1\alpha} e^{-i\theta_2/2} & \frac{i}{\sqrt{2}} \left(Y'_{1\alpha}-Y'_{2\alpha}\right) e^{-i\theta_2/2} \\[8pt] 
Y_{13}^*Y_{1\alpha} v_\xi/M_{33} &Y_{3\alpha} 
\end{pmatrix}.\quad\,\,\,
\label{eq:couplings}
\end{eqnarray}
Unlike in the previous section, there is a phase difference in the Yukawa couplings between $\phi_1$ and leptons, namely $Y_{1\alpha}e^{-i\theta_1/2}/\sqrt{2}$ and $Y_{1\alpha}e^{-i\theta_2/2}/\sqrt{2}$, and the asymmetry is proportional to $\sin(\theta_2 - \theta_1)$. If $\theta_2 = \theta_1$, the asymmetry vanishes as proved at the end of Sec.\ref{sec:Ib_pseu_asym}.

\begin{figure}[t!]
\begin{center}
\subfigure[]{ \includegraphics[height=0.15\textwidth]{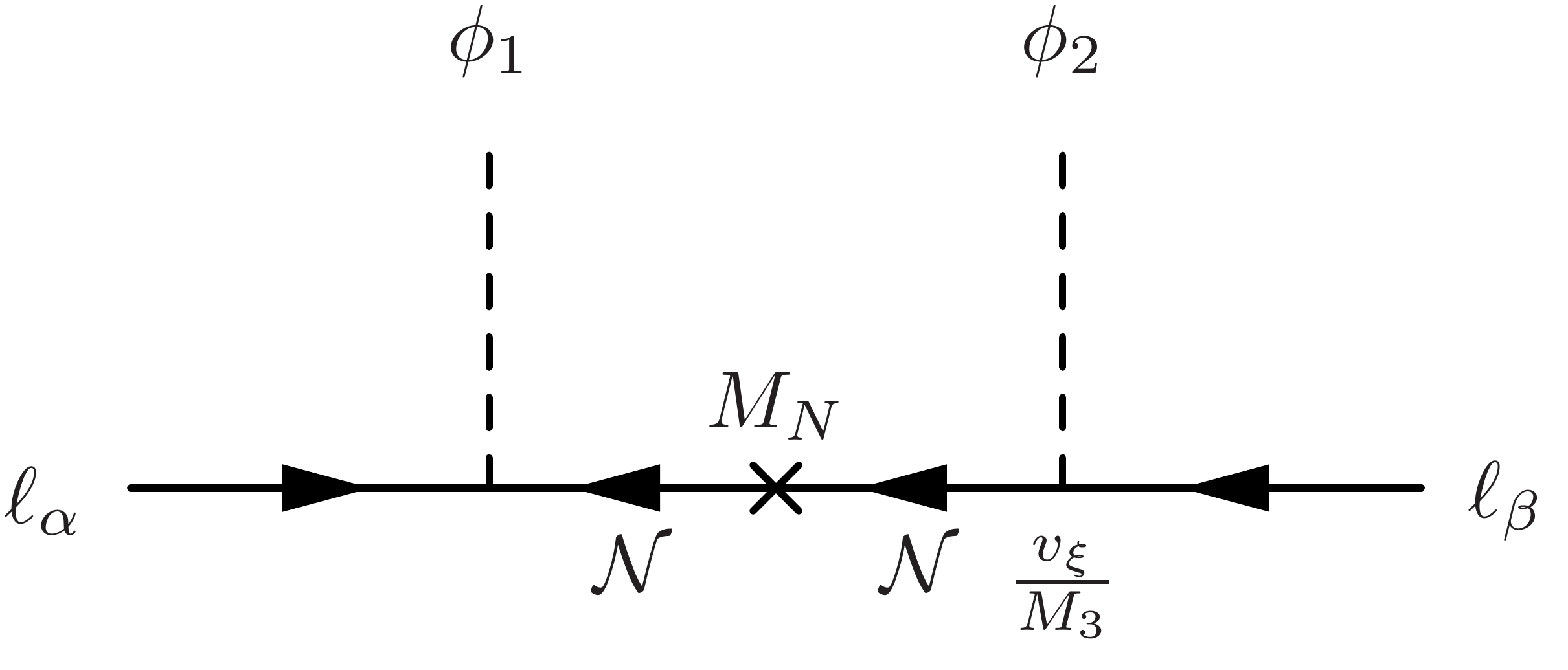} \label{fig:NMexdirac}}
\subfigure[]{ \includegraphics[height=0.15\textwidth]{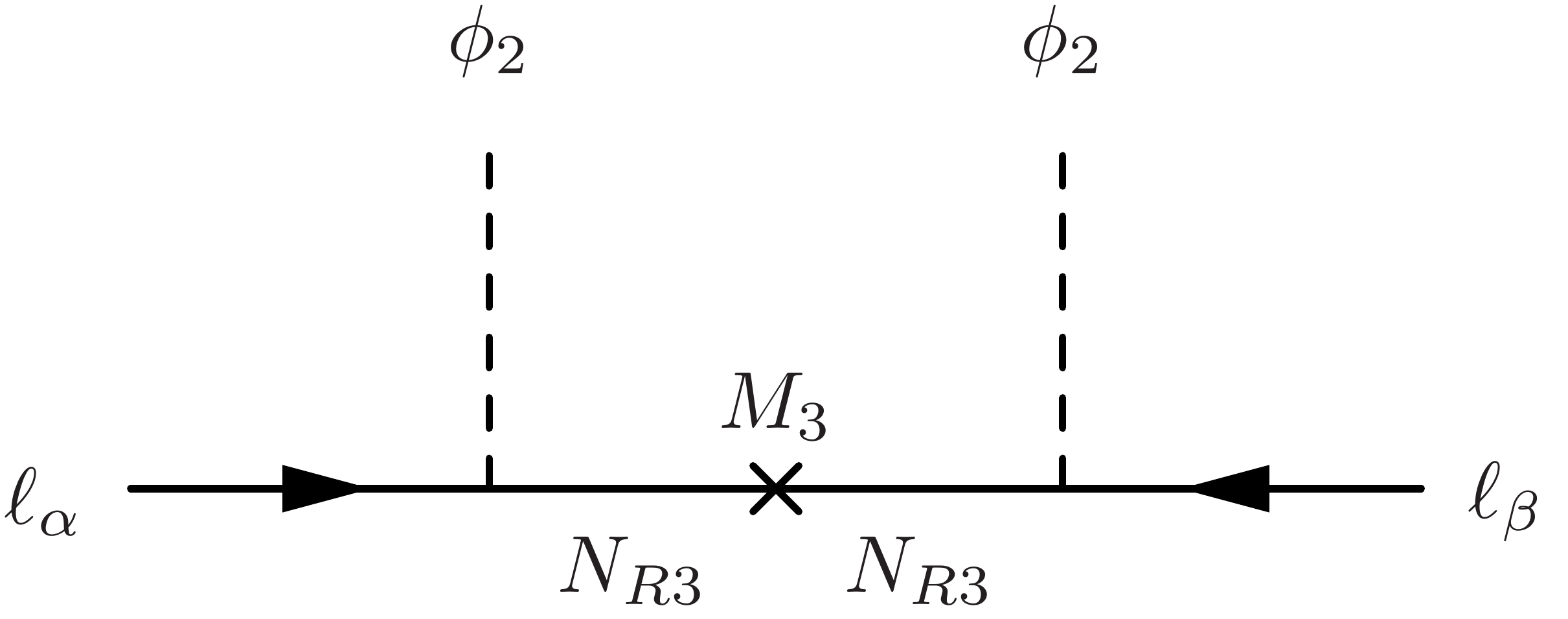} \label{fig:NMexmaj}}
\caption{\label{fig:NMIb_ex} Neutrino mass in the extended type Ib seesaw model.}
\end{center}
\end{figure}

Following the procedure in \cite{Broncano:2002rw}, we integrate out the heavy neutrino to generate a set of effective operators, which leads to an effective field theory for the low energy phenomenology. The dimension-five effective operators are Weinberg-type operators involving two different Higgs doublets which looks like a simple extension of the minimal type Ib seesaw model \cite{Hernandez-Garcia:2019uof}
\begin{eqnarray}
\delta \mathcal{L}^{d=5} &=& 
-\frac{1}{2M}\left[ Y_{1\alpha}^*(Y'_{1\beta})^*\left(e^{i\theta_1/2} - e^{i\theta_2/2}\right) \right.\nonumber\\&&\left.+ Y_{1\alpha}^* (Y'_{2\beta})^*\left(e^{i\theta_1/2} + e^{i\theta_2/2}\right) \right] \overline{\ell^c}_\alpha{\phi}_1^*{\phi}_2^\dagger \ell_\beta \nonumber\\ &&
+ \frac{1}{M_{33}}Y_{3\alpha}^*Y_{3\beta}^*\, \overline{\ell^c}_\alpha\phi_2^*\phi_2^\dagger \ell_\beta + {\rm h.c.},
\label{eq:d5op_ex_full}
\end{eqnarray}
where $\ell_\alpha$ are the lepton doublets and $\phi_i$ are the Higgs doublets. Since $\theta_1$ and $\theta_2$ are both small, the first term in the first line of Eq.\eqref{eq:d5op_ex_full} is neglectable and at leading order the operators are
\begin{eqnarray}
\delta \mathcal{L}^{d=5} &=&
-\frac{1}{M}Y_{1\alpha}^*(Y'_{2\beta})^* \overline{\ell^c}_\alpha{\phi}_1^*{\phi}_2^\dagger \ell_\beta \nonumber\\&&+ \frac{1}{M_{33}}Y_{3\alpha}^*Y_{3\beta}^*\, \overline{\ell^c}_\alpha\phi_2^*\phi_2^\dagger \ell_\beta + {\rm h.c.}.
\label{eq:d5op_ex}
\end{eqnarray}
This result can also be explained in the original basis where $N_{R1}$ and $N_{R2}$ form a pseudo-Dirac fermion. The two terms in the Weinberg operator Eq.\eqref{eq:d5op_ex} correspond to Fig.\ref{fig:NMexdirac} and Fig.\ref{fig:NMexmaj} and are both of order $1/M_{33}$ since the Yukawa interaction of $N_{R2}$ is effective. 
Since we are interested in the case where the neutrino mass originates from type Ib seesaw mechanism, it is assumed that
\begin{eqnarray}
\left|Y_{1\alpha}^*Y_{23}^*\right|v_\xi \gg \left|Y_{3\alpha}^*\right|M \label{eq:Ibmass_con}
\end{eqnarray}
so that the first term in Eq.\eqref{eq:d5op_ex} is dominant. As before, the Yukawa couplings in the normal hierarchy (NH), read
\begin{eqnarray}
Y_{1\alpha} &=&\dfrac{Y_1}{\sqrt{2}}\left( \sqrt{1+\rho} \left(U_\text{PMNS}\right)_{\alpha 3}-\sqrt{1-\rho} \left(U_\text{PMNS}\right)_{\alpha 2}\right), \nonumber\\
\label{eq:Y1_NH_ex}\\
Y'_{2\alpha} &=& \dfrac{Y_2}{\sqrt{2}}\left( \sqrt{1+\rho} \left(U_\text{PMNS}\right)_{\alpha 3}+\sqrt{1-\rho} \left(U_\text{PMNS}\right)_{\alpha 2}\right), \nonumber\\
\label{eq:Y2_NH_ex}
\end{eqnarray}
and the couplings satisfy 
\begin{eqnarray}
\sum_{\alpha}Y_{1\alpha}^* Y_{1\alpha} = Y_1^2, &\quad& 
\sum_{\alpha}(Y'_{2\alpha})^* Y'_{2\alpha} = Y_2^2,\nonumber\\
\sum_{\alpha}Y_{1\alpha}^* Y'_{2\alpha} &=&  \rho Y_1Y_2.
\label{eq:relY_ex}
\end{eqnarray} 
The real quantities $Y_1$ and $Y_2$ follow the relation Eq.\eqref{eq:rel2}.

\subsection{CP asymmetry for resonant leptogenesis \label{sec:Ib_ex_asym}}

With a mass splitting in the RH neutrino mass, the decay rates of the RH neutrinos are 
\begin{eqnarray}
\Gamma^{\rm tree}_{n_1\ell}&=&\Gamma^{\rm tree}_{n_1\overline{\ell}}\simeq\frac{Y_1^2}{32\pi}M_1 ,\\
\Gamma^{\rm tree}_{n_2\ell}&=&\Gamma^{\rm tree}_{n_2\overline{\ell}}\simeq\frac{Y_1^2}{32\pi}M_2 ,
\label{eq:dec_tree_ex}
\end{eqnarray} 
both of which have the same leading order limit $Y_1^2M/32\pi$. In the resonant leptogenesis regime, the CP asymmetry is dominantly produced from the decay of heavy neutrinos with self-energy. The corresponding contributions are given by 
\begin{eqnarray}
\left(\epsilon_{n_1}^{\rm wave}\right)_{k\alpha}&\equiv&
\frac{\Gamma_{n_1\rightarrow \ell_\alpha\phi_k}^{\rm wave} - \Gamma_{n_1\rightarrow\overline{\ell}_\alpha\phi_k^\dagger}^{\rm wave}}{\Gamma_{n_1\ell} + \Gamma_{n_1\overline{\ell}}} 
\nonumber\\&=& \frac{-\Im\left[ \left(y_{1k\alpha}y_{2k\alpha}^*\right) \sum_{l,\beta} \left(y_{1l\beta}y_{2l\beta}^*\right)\right]}{\sum_{k,\alpha} \left(y_{1k\alpha}y_{1k\alpha}^*\right) \sum_{l,\beta} \left(y_{2l\beta}y_{2l\beta}^*\right)} \times
\nonumber\\&&\frac{2 \Delta M_{12} \Gamma_{n_2}}{4\left(\Delta M_{12}\right)^2 + \Gamma_{n_2}^2},
\label{eq:epswn_ex_1}\\
\left(\epsilon_{n_2}^{\rm wave}\right)_{k\alpha}&\equiv&
\frac{\Gamma_{n_2\rightarrow \ell_\alpha\phi_k}^{\rm wave} - \Gamma_{n_2\rightarrow\overline{\ell}_\alpha\phi_k^\dagger}^{\rm wave}}{\Gamma_{n_2\ell} + \Gamma_{n_2\overline{\ell}}} 
\nonumber\\&=& \frac{-\Im\left[\left(y_{1k\alpha}y_{2k\alpha}^*\right) \sum_{l,\beta} \left(y_{1l\beta}y_{2l\beta}^*\right) \right]}{\sum_{k,\alpha} \left(y_{1k\alpha}y_{1k\alpha}^*\right) \sum_{l,\beta} \left(y_{2l\beta}y_{2l\beta}^*\right)} \times
\nonumber\\&&\frac{2 \Delta M_{12} \Gamma_{n_1}}{4\left(\Delta M_{12}\right)^2 + \Gamma_{n_1}^2},
\label{eq:epswn_ex_2}
\end{eqnarray} 
where $\Delta M_{12} = M_2 - M_1$. The resonance condition is $\Delta M_{12} \simeq \Gamma_{n_i}/2$. 
Using Eq.\eqref{eq:couplings}, one can get
\begin{eqnarray}
y_{11\beta}y_{21\beta}^* &=&\frac{i}{2}Y_{1\beta}Y_{1\beta}^*e^{i(\theta_2-\theta_1)/2},\,\\
y_{12\beta}y_{22\beta}^* &=&\frac{i}{2}\left(Y'_{1\beta}+Y'_{2\beta}\right)\left(Y'_{1\beta}-Y'_{2\beta}\right)^*e^{i(\theta_2-\theta_1)/2} .\quad 
\end{eqnarray} 
Since the couplings $Y'_{1\alpha}$ and $Y'_{2\alpha}$ are suppressed by $v_\xi/M_{33}$, there is 
\begin{eqnarray}
\sum_{l,\beta} \left(y_{1l\beta}y_{2l\beta}^*\right) &\simeq& \sum_{\beta} \frac{i}{2}Y_{1\beta}Y_{1\beta}^*e^{i(\theta_2-\theta_1)/2} \nonumber\\ &\simeq& \frac{i}{2}Y_1^2e^{i(\theta_2-\theta_1)/2}
\end{eqnarray} 
and the leading order contribution is 
\begin{eqnarray}
\left(\epsilon_{n_1}^{\rm wave}\right)_{1\alpha}&=& \frac{Y_{1\alpha}Y_{1\alpha}^* \sin(\theta_2-\theta_1)}{Y_1^2} \frac{2 \Delta M_{12} \Gamma_{n_2}}{4\left(\Delta M_{12}\right)^2 + \Gamma_{n_2}^2},
\label{eq:epswn1_lead}\\
\left(\epsilon_{n_2}^{\rm wave}\right)_{1\alpha}&=& \frac{Y_{1\alpha}Y_{1\alpha}^* \sin(\theta_2-\theta_1)}{Y_1^2} \frac{2 \Delta M_{12} \Gamma_{n_1}}{4\left(\Delta M_{12}\right)^2 + \Gamma_{n_1}^2}.\quad\quad
\label{eq:epswn2_lead}
\end{eqnarray} 
The production of asymmetry relies on the difference in angle $\theta_i$, which vanishes in the previous model. With Eq.\eqref{eq:masses_re} and Eq.\eqref{eq:dec_tree_ex}, it can be derived that the resonant condition \footnote{The resonant condition can be affected by the thermal effects \cite{Granelli:2020ysj}. However, as will be shown later, the parameter space allowed by various constraints is not sensitive to the resonant condition. } is satisfied when 
\begin{eqnarray}
\frac{\Re(M_{13}^2 + M_{23}^2)}{M M_{33}}\simeq \frac{Y_1^2}{32\pi} \label{eq:res_con}
\end{eqnarray} 
and the resonant asymmetry under such condition is 
\begin{eqnarray}
\epsilon_\alpha &=& \left(\epsilon_{n_1}^{\rm wave}\right)_{1\alpha} = \left(\epsilon_{n_2}^{\rm wave}\right)_{1\alpha} 
\nonumber\\ &=& \frac{Y_{1\alpha}Y_{1\alpha}^* \sin(\theta_2-\theta_1)}{2Y_1^2} 
\simeq \frac{Y_{1\alpha}Y_{1\alpha}^*}{2Y_1^2}\frac{\Im\left(M_{13}^2+M_{23}^2\right)}{MM_{33}}.\label{eq:epswn_res}\quad \quad
\end{eqnarray} 


\subsection{Boltzmann equations and the approximate solution\label{sec:Ib_ex_BE}}
As we aim to study GeV heavy neutrinos, the asymmetry needs to be produced at $z\sim O(0.01)$ since the EW sphaleron decouples at $T_{sph}=131$ GeV \cite{DOnofrio:2014rug}. However, in the type Ib seesaw model, the inverse decay is too strong that it decouples at $z \sim 20$. In order to have the inverse decay decoupling temperature higher than the sphaleron decoupling temperature \footnote{The sphaleron decoupling temperature is computed for the SM instead of 2HDM, but we do not expect too much difference in 2HDM.}, the heavy neutrino mass has to be above 4 TeV. For sub-TeV scale heavy neutrinos, the lepton asymmetry can still be produced before the EW sphaleron decouples. As the inverse decay of heavy neutrinos is strong, the heavy neutrinos can be easy produced in the early univese and thus a thermal initial abundance of the heavy neutrinos is likely to exist. If the EW sphaleron decouples before the comoving density of heavy neutrinos is Boltzmann suppressed ($z\ll1$), enough amount of asymmetry can survive after the sphaleron decouples. 

In such a scenario, the lepton asymmetry is produced when the RH neutrinos are relativistic and therefore the thermal effect is important \cite{Biondini:2017rpb,Granelli:2020ysj}. In fact, the thermal mass of the Higgs doublets can be large enough so that it can decay into RH neutrinos and leptons at $z<1$ \cite{Hambye:2016sby}. After taking the thermal effect into account, the asymmetry factors in Eq.\eqref{eq:epswn1_lead} and Eq.\eqref{eq:epswn2_lead} become \cite{Hambye:2016sby,Granelli:2020ysj}
\begin{eqnarray}
\left(\epsilon_{n_1}^{\rm wave}\right)_{1\alpha}&=& \frac{Y_{1\alpha}Y_{1\alpha}^* \sin(\theta_2-\theta_1)}{Y_1^2} \times
\nonumber\\&& \frac{2 \Delta M_{12} \Gamma_{n_2}\zeta(z)}{4\left(\Delta M_{12} + \Delta M_{12}^T(z)\right)^2 + \Gamma_{n_2}^2\zeta^2(z)},
\label{eq:epswn1_thermal}\\
\left(\epsilon_{n_2}^{\rm wave}\right)_{1\alpha}&=& \frac{Y_{1\alpha}Y_{1\alpha}^* \sin(\theta_2-\theta_1)}{Y_1^2} \times
\nonumber\\&& \frac{2 \Delta M_{12} \Gamma_{n_1}\zeta(z)}{4\left(\Delta M_{12} + \Delta M_{12}^T(z)\right)^2 + \Gamma_{n_1}^2\zeta^2(z)}. \quad\quad
\label{eq:epswn2_thermal}
\end{eqnarray} 
$\Delta M_{12}^T(z)$ is the variation in $\Delta M_{12}$ due to thermal effect, given by
\begin{eqnarray}
\Delta M_{12}^T(z)\simeq\frac{Y_1^2 M}{32z^2}.
\end{eqnarray} 
The factor $\zeta(z)$ quantifies the temperature dependence in the RH neutrino decay rates, which behaves as a constant $\zeta_0$ at high temperature \cite{Granelli:2020ysj}. The exact value of $\zeta_0$, as well as the threshold temperature for Higgs decay, depends on the details in phase transition and stability of the Higgs potential. Since detailed discussion about the Higgs potential is beyond the scope of this paper, we assume the leptogenesis is realised through the Higgs decay for 1-100 GeV neutrinos and choose several benchmark values for $\zeta_0$ to show how the leptogenesis can be affected. 

With the thermal effects, the resonant condition in Eq.\eqref{eq:res_con} is modified into
\begin{eqnarray}
\frac{\Re(M_{13}^2 + M_{23}^2)}{M M_{33}}\simeq \frac{Y_1^2}{32\pi} \sqrt{\zeta_0^2 + \frac{\pi^2}{z^4}}\label{eq:res_con_TE}
\end{eqnarray} 
and the resonant asymmetry in Eq.\eqref{eq:epswn_res} also changes into
\begin{eqnarray}
\epsilon_\alpha (z)
&\simeq& \frac{Y_{1\alpha}Y_{1\alpha}^*}{2Y_1^2}\frac{\Im\left(M_{13}^2+M_{23}^2\right)}{MM_{33}}\frac{\sqrt{\zeta_0^2 + \tfrac{\pi^2}{z^4}}-\tfrac{\pi}{z^2}}{\zeta_0} \nonumber\\
&\simeq &\epsilon_\alpha 
\begin{dcases}
1 & \zeta_0 > \tfrac{\pi}{z^2},\\
\tfrac{\zeta_0z^2}{2\pi} & \zeta_0 < \tfrac{\pi}{z^2}.
\end{dcases}\quad \quad
\end{eqnarray} 
Let $z_\zeta=\sqrt{\pi/ \zeta_0}$. The typical value of the thermal factor $\zeta_0$ is of order 10 \cite{Granelli:2020ysj}, which indicates that the critical $z_\zeta$ is of the same order as $z_{sph}=M/T_{sph}$ for 1-100 GeV heavy neutrinos.

The Boltzmann equations can be written as \cite{Buchmuller:2004nz,Abada:2006ea,Granelli:2020ysj} 
\begin{eqnarray}
\frac{dY_{n_i}}{dz} &=& - \mathcal{C}_1 \kappa_i \left(Y_{n_i} -Y_{n_i}^{\rm eq}\right) ,\\
\frac{dY_{\Delta_{\alpha}}}{dz} &=& \sum_i -\epsilon_{i\alpha}\mathcal{C}_1 \kappa_i \left(Y_{n_i} -Y_{n_i}^{\rm eq}\right) +\mathcal{C}_2 \kappa_{i\alpha} A_{\alpha\beta}Y_{\Delta_{\beta}},\quad\quad
\end{eqnarray}
where $\epsilon_{i\alpha}=\left(\epsilon_{n_i}^{\rm wave}\right)_{1\alpha}$, $\kappa_i=\tilde{m}_i/m^*_i$ and $\kappa_{i\alpha}=\tilde{m}_{i\alpha}/m^*_i$. The coefficients $\mathcal{C}_1$ and $\mathcal{C}_2$ are determined by 
\begin{eqnarray}
\mathcal{C}_1 &=& \frac{z}{\s Y^{\rm eq}_{n_i}} \frac{ \gamma_{1\rightarrow2}^{n_i} + \gamma_{2\rightarrow2}^{n_iA}}{\Gamma_{n_i}},\quad\quad\\
\mathcal{C}_2 &=& \frac{z}{\s Y^{\rm eq}_{n_i}} \frac{\left( \gamma_{W,1\rightarrow2}^{n_i} + \gamma_{W,2\rightarrow2}^{n_iA} \right)_\alpha}{\left(\Gamma_{n_i}\right)_\alpha}.\quad\quad
\label{eq:BE_ni_sim_ex}
\end{eqnarray} 
At $z\gg1$, both $\gamma_{1\rightarrow2}^{n_i}$ and $\gamma_{2\rightarrow2}^{n_iA}$ are proportional to $T^4$ \cite{Besak:2012qm} . The tempereture dependence of $\mathcal{C}_1$ and $\mathcal{C}_2$ is cancelled, thus they can be treated as constants in the Boltzmann equation. Here, the $2\rightarrow2$ scattering involve top quark is neglectable because the interactions between $\phi_2$ and the quasi-degenerate RH neutrinos are suppressed by $1/M_{33}$ while the top quark only couples to $\phi_2$. The bottom quark scattering process may contribute instead as the bottom quark mass is sourced from $\phi_1$ with the smaller VEV in the 2HDM. The total decay rates of $2\rightarrow 2$ scattering proceses follow \cite{Besak:2012qm} 
\begin{eqnarray}
\gamma_{2\rightarrow2}^{n_iA}&=&\frac{Y_1^2 T^4}{3072\pi} \left(3g^2 +{g'}^2\right)\left[3.17-\ln\left(3g^2 +{g'}^2\right)\right],\quad\quad\\
\gamma_{2\rightarrow2}^{n_ib}&=&\frac{Y_1^2 T^4}{3072\pi}\left(Y^d_{33}\right)^2 \times 2.52\,.
\end{eqnarray}
The bottom quark Yukawa coupling $Y^d_{33}$ is around $0.024 \cot\beta$, while $\sqrt{3g^2 +{g'}^2}\simeq1.2$. For large $\tan\beta$, the bottom quark scattering process can also make significant contribution to the Boltzmann equations. Here, for simplicity, the contribution from bottom quark scattering is made neglectable through adjusting the value of $\tan\beta$.

Although these coefficients have some temperature-dependence, the dependence is very weak when $z<1$ \cite{Besak:2012qm}, especially after taking soft gauge interactions into account, and hence they can be treated as constants equal to their respective average values in the interested interval of temperature \cite{Granelli:2020ysj}. By defining $\Delta_{n_i} \equiv Y_{n_i} -Y_{n_i}^{\rm eq}$, the Boltzmann equations can be turned into
\begin{eqnarray}
\frac{d\Delta_{n_i}}{dz} &=& - \mathcal{C}_1 \kappa_i\Delta_{n_i} - \frac{dY_{n_i}^{\rm eq}}{dz} ,\\
\frac{dY_{\Delta_{\alpha}}}{dz} &=& \sum_i \epsilon_{i\alpha}\mathcal{C}_1 \kappa_i \Delta_{n_i} + \mathcal{C}_2 \kappa_{i\alpha} A_{\alpha\beta} Y_{\Delta_{\beta}}.
\end{eqnarray}
At high temperature ($z\ll1$), $d\Delta_{n_i}/dz$ is neglectable and the approximate solution of $\Delta_{n_i}$ is
\begin{eqnarray}
\Delta_{n_i} &\simeq& \frac{z}{2\,\mathcal{C}_1 \kappa_i} Y_{n_i}^{\rm eq}\,.
\end{eqnarray}
The comoving density of $B-L_\alpha$ is then
\begin{widetext}
\begin{eqnarray}
Y_{\Delta_{\alpha}} &\simeq&\sum_i \frac{\epsilon_\alpha Y_{n_i}^{\rm eq}}{2}
\begin{dcases}
\frac{\zeta_0}{2\pi} \frac{6 e^{-h_\alpha z} - 6 + 6h_\alpha z - 3 h_\alpha^2 z^2 + h_\alpha^3 z^3}{h_\alpha^4} & z < z_\zeta, \\
\frac{\zeta_0}{2\pi} \frac{6 e^{-h_\alpha z_\zeta} - 6 + 6h_\alpha z_\zeta - 3 h_\alpha^2 z_\zeta^2 + h_\alpha^3 z_\zeta^3}{h_\alpha^4\, e^{h_\alpha(z-z_\zeta)}}  + \frac{e^{-h_\alpha (z-z_\zeta)}(1- h_\alpha z_\zeta) - 1+ h_\alpha z}{h_\alpha^2} & z> z_\zeta,
\end{dcases}\quad\quad
\end{eqnarray}
\end{widetext}
where $h_\alpha=\mathcal{C}_2 \kappa_{i\alpha} |A_{\alpha\alpha}|$. The off-diagonal elements of the $A$-matrix are ignored. The final $B-L$ asymmetry transferred into baryon asymmetry is determined by $Y_{\Delta_{\alpha}}$ at $T_{sph}$, at which temperature there is
\begin{eqnarray}
h_\alpha\, z_{sph} \simeq 1.3\times10^{13} |Y_{1\alpha}|^2.
\end{eqnarray}
where the coefficient $\mathcal{C}_2$ is roughly taken to be $0.1$ \cite{Abada:2006ea,Besak:2012qm,Granelli:2020ysj}. For $M\in (1,100)$ GeV, the product of $Y_1$ and $Y_2$ is roughly between $O(10^{-12})$ and $O(10^{-14})$. As $Y_1\gg Y_2$, the quantity $h_\alpha\, z_{sph}$ is far larger than 1 and the approximate solution of $Y_{\Delta_{\alpha}}$ at $T_{sph}$ can be reduced to 
\begin{eqnarray}
Y_{\Delta_{\alpha}} &\simeq& \epsilon_\alpha Y_{n}^{\rm eq}
\begin{dcases} 
\frac{\zeta_0}{2\pi} \frac{z_{sph}}{h_\alpha} & z_{sph} < z_\zeta,\\
\frac{\zeta_0}{2\pi} \frac{z_\zeta}{h_\alpha} e^{h_\alpha(z_\zeta - z_{sph})} + \frac{z_{sph}}{h_\alpha} & z_{sph} > z_\zeta,
\end{dcases}\quad\quad
\end{eqnarray}
Unless $(z_{sph} - z_\zeta) \sim O(h_\alpha^{-1})$, the expression of $Y_{\Delta_{\alpha}}$ for $z_{sph} > z_\zeta$ can be reduced into $z_{sph}/h_\alpha$. The constraint from observed baryon asymmetry is then 
\begin{eqnarray}
\frac{\Im\left(M_{13}^2+M_{23}^2\right)}{M M_{33}} \frac{z_{sph}^2}{Y_1^2} \simeq 5.7 \times 10^5 \Theta(z_\zeta),
\label{eq:con_fin_light}
\end{eqnarray}
where
 \begin{eqnarray}
\Theta(z_\zeta) = 
\begin{dcases}
2\pi/\zeta_0 & z_{sph} < z_\zeta,\\
1 & z_{sph} > z_\zeta.
\end{dcases}
\end{eqnarray}
In the case of the SM limit, $\zeta_0$ is around 23 \cite{Granelli:2020ysj} and $z_\zeta \simeq 0.37$. The condition $z_{sph} < z_\zeta$ is satisfied when the RH neutrino is lighter than 48 GeV and the thermal effect varies the constraint from requiring the correct baryon asymmetry in Eq.\eqref{eq:con_fin_light} by a factor $0.27$.

\subsection{Results and Discussion}

In the original seesaw Lagrangian Eq.\eqref{eq:lg_Ibex}, there are 7 free parameters: 3 mass scales $M$, $M_{33}$ and $v_\xi$, 2 seesaw Yukawa couplings $Y_{1\alpha}$ and $Y_{3\alpha}$, 2 Yukawa couplings between RH neutrinos $Y_{13}$ and $Y_{23}$. Although $Y_{3\alpha}$ does not appear in the type Ib seesaw Weinberg operator directly, it can still be expressed by analogue to Eq.\eqref{eq:Y2_NH_ex} as 
\begin{eqnarray}
Y_{3\alpha} &=& \dfrac{Y_3}{\sqrt{2}}\left( \sqrt{1+\rho} \left(U_\text{PMNS}\right)_{\alpha 3}+\sqrt{1-\rho} \left(U_\text{PMNS}\right)_{\alpha 2}\right)\nonumber\\&&\times e^{-i\arg(M_{23})}, 
\label{eq:Y3_NH_ex}
\end{eqnarray}
since $Y'_{2\alpha}= 2M_{23}Y_{3\alpha}/M_{33}$. Then a sufficient condition for Eq.\eqref{eq:Ibmass_con} can be obtained by minimise the fraction of normilised component vectors
\begin{eqnarray}
\frac{Y_1|M_{23}|^2}{Y_2 M M_{33}}\gg 2.5.
\label{eq:Ibmass_con_2}
\end{eqnarray}
Notice that the parameters $M_{13}$, $M_{23}$, $M_{33}$ always appear as $M_{13}^2/M_{33}$ and $M_{23}^2/M_{33}$ in the constraints and assumptions so far. Therefore we defined dimensionless quantities $R_1\equiv M_{13}^2/MM_{33}$ and $R_2\equiv M_{23}^2/MM_{33}$ for convenience.\footnote{The scale of $v_\xi$ and $M_{33}$ should be properly chosen to keep the couplings $Y_{13}$ and $Y_{23}$ below their perturbativity limit.} As the higher energy scales $v_\xi$ and $M_{33}$ are hard to be constrained by experiments, $R_1$ and $R_2$ can be considered as free parameters. Eq.\eqref{eq:res_con} and Eq.\eqref{eq:con_fin_light} can be combined into one equation as 
\begin{eqnarray}
R_1+R_2 &\simeq& \frac{Y_1^2}{32\pi} \left(1 + 5.7\times10^7 z_{sph}^{-2}\Theta(z_\zeta)\,i\right).
\end{eqnarray}
If the pseudo-Dirac RH neutrino is not superheavy, the sum $R_1 + R_2$ is almost completely imaginary. With the aim of studying testable neutrino phenomenology, the neutrino mass is considered to be below TeV scale, in which case the real part of $R_1 + R_2$ is neglectable. Then the constraints and assumptions can then be summarised as 
\begin{eqnarray}
|R_2|Y_1/Y_2 &\gg& 2.5 .\label{eq:cons_Ib_dom}\\
|R_1|,|R_2| &\ll& 1 ,\label{eq:cons_pdirac}\\
\frac{Y_1 Y_2 v\sin{\beta}\cos{\beta}}{M} &\simeq& 2.4 \times 10^{-13},\label{eq:cons_osc}\\
R_1+R_2 &\simeq& 5.7\times10^5\,Y_1^2 z_{sph}^{-2}\Theta(z_\zeta)i,\label{eq:cons_lep}\quad\quad
\end{eqnarray}
Eq.\eqref{eq:cons_Ib_dom} is the condition for type Ib dominant light neutrino mass; Eq.\eqref{eq:cons_pdirac} is assumed for the pseudo-Dirac structure of RH neutrinos $N_{R1}$ and $N_{R2}$; Eq.\eqref{eq:cons_osc} is required for generating the correct neutrino mass and mixing in the type Ib seesaw model; Eq.\eqref{eq:cons_lep} is the constraint from the observed baryogenesis. Although Eq.\eqref{eq:cons_Ib_dom} and Eq.\eqref{eq:cons_pdirac} are not experimental or observational constraints, the type Ib structure relies on these relations. If these conditions are not satisfied, the neutrino mass would be generated by type Ia seesaw mechanism instead and the seesaw couplings and RH neutrino mass receive different constraints from the oscillation data and observed baryogenesis. Therefore Eq.\eqref{eq:cons_Ib_dom} and Eq.\eqref{eq:cons_pdirac} are required for our interest in the type Ib seesaw mechanism and testable results. On the other hand, it is worth noticing that the resonant condition in Eq.\eqref{eq:res_con} does not play any role in constraining the seesaw Yukawa coupling as the other constraints Eq.\eqref{eq:cons_Ib_dom} and Eq.\eqref{eq:cons_pdirac} only depend on $|R_1|$ and $|R_2|$ which are determined by the imaginary part of $R_1$ and $R_2$ while the $R_1$ and $R_2$ themselves are determined by higher scale quantities beyond testability.

Having assumed Eq.\eqref{eq:cons_Ib_dom} and Eq.\eqref{eq:cons_pdirac}, it
can be deduced that these quantities play critical roles in constraining the parameter space. Without these equations, the only constraint on the seesaw parameters is Eq.\eqref{eq:cons_osc}, as $R_1$ and $R_2$ are completely free in Eq.\eqref{eq:cons_lep}. To see how Eq.\eqref{eq:cons_Ib_dom} and Eq.\eqref{eq:cons_pdirac} constrain the parameter space, we consider three different cases: $|R_1|\ll |R_2|$, $|R_1| \gg |R_2|$ and $|R_1| \simeq |R_2|$. If $|R_1|\ll |R_2|$, there is 
\begin{eqnarray}
|R_2| \simeq 5.7\times10^5\,Y_1^2 z_{sph}^{-2} \Theta(z_\zeta) , \label{eq:Rcase1}
\end{eqnarray}
and Eq.\eqref{eq:cons_Ib_dom} and Eq.\eqref{eq:cons_pdirac} imply 
\begin{widetext}
\begin{eqnarray}
1.3 \times10^{-3} \frac{z_{sph}}{\sqrt{\Theta(z_\zeta)}} \gg Y_1 \gg 3.2 \times 10^{-5}\sqrt{z_{sph}}\left(\frac{M}{v\sin{\beta}\cos{\beta}}\right)^{1/4} \Theta(z_\zeta)^{-1/4} .
\label{eq:Rcase1_range}
\end{eqnarray}
\end{widetext}
Notice that the condition Eq.\eqref{eq:cons_Ib_dom} is very sensitive to changes in $Y_1$ in this case as the left side is proportional to $Y_1^4$. If $|R_1| \gg |R_2|$, then there is 
\begin{eqnarray}
|R_1| \simeq 5.7\times10^5\,Y_1^2 z_{sph}^{-2} \Theta(z_\zeta) \,
\end{eqnarray}
instead of Eq.\eqref{eq:Rcase1}. The upper bound of $Y_1$ is the same as in the first case since $|R_1|$ and $|R_2|$ receive the same constraint in Eq.\eqref{eq:cons_pdirac}, while the lower bound of $Y_1$ is larger than the one of the first case as implied by Eq.\eqref{eq:cons_Ib_dom}. Therefore the allowed range of $Y_1$ in this case is smaller than the one in the case $|R_1|\ll |R_2|$. If $|R_1| \simeq |R_2|$, then there is 
\begin{eqnarray}
|R_2|\simeq |R_1| \gg 5.7\times10^5\,Y_1^2 z_{sph}^{-2} \Theta(z_\zeta). \label{eq:Rcase3}
\end{eqnarray}
Eq.\eqref{eq:Rcase3} together with Eq.\eqref{eq:cons_Ib_dom} and Eq.\eqref{eq:cons_pdirac} imply 
\begin{widetext}
\begin{eqnarray}
1.3 \times10^{-3} z_{sph} \sqrt{\frac{|R_1|}{\Theta(z_\zeta)}}\gg Y_1 \gg 5.7 \times 10^{-7}\sqrt{\frac{z_{sph}}{\sin{\beta}\cos{\beta}|R_1|}}.\label{eq:Rcase3_range}
\end{eqnarray}
\end{widetext}
To have a solution in Eq.\eqref{eq:Rcase3_range} when $\tan\beta=10$, $|R_1|$ has to be at least of order $0.01/\sqrt{z_{sph}}$ while also much smaller than 1, which means the heavy neutrino cannot be much below GeV scale. The largest allowed range of $Y_1$ as a combination of Eq.\eqref{eq:Rcase1_range} and Eq.\eqref{eq:Rcase3_range} can be obtained by taking $|R_1|\simeq 0.1$
\begin{eqnarray}
1.3 \times10^{-3} \frac{z_{sph}}{\sqrt{\Theta(z_\zeta)}} \gg Y_1 \gg 1.8 \times 10^{-6}\sqrt{\frac{z_{sph}}{\sin{\beta}\cos{\beta}}}.\label{eq:Rcall_range}\quad\quad
\end{eqnarray}
Some benchmark cases for 1, 10 and 100 GeV heavy neutrino when $z_{sph} > z_\zeta$ are shown in Tab.\ref{tab:bench}.
\begin{table}[t!]
\centering
\begin{tabular}{|c|c|c|c|c|c|}
\hline & & & &  \\ [-1em]
$M$ [GeV] & $Y_1$ & $R_1 $ & $R_2$ & $|R_2|Y_1/Y_2$ \\[2pt] 
\hline & & & &  \\ [-1em]
$1$ & $3\times10^{-6}$ & 0 & $0.088i$ & $80$ \\ 
\hline & & & &  \\ [-1em]
$1$ & $2\times10^{-6}$ & 0 & $0.039i$ & $16$ \\ 
\hline & & & &  \\ [-1em]
$1$ & $2\times10^{-6}$ & $0.1$ & $-0.1+ 0.039i$ & $44$ \\ 
\hline & & & &  \\ [-1em]
$10$ & $3\times10^{-5}$ & 0 & $0.088i$ & $8.0\times10^2$ \\ 
\hline & & & &  \\ [-1em]
$10$ & $2\times10^{-5}$ & 0 & $0.039i$ & $1.6\times10^2$ \\ 
\hline & & & &  \\ [-1em]
$10$ & $5\times10^{-6}$ & $0.1$ & $-0.1 + 0.0024i$ & $25$ \\ 
\hline & & & &  \\ [-1em]
$100$ & $3\times10^{-4}$ & 0 & $0.088i$ & $8.0\times10^3$ \\ 
\hline & & & &  \\ [-1em]
$100$ & $1\times10^{-4}$ & 0 & $0.0098i$ & $99$ \\ 
\hline & & & &  \\ [-1em]
$100$ & $2\times10^{-5}$ & $0.1$ & $-0.1 + 0.0004i$ & $41$ \\ 
\hline 
\end{tabular}
\caption{\label{tab:bench}Benchmark values of the free parameters for sub-TeV scale heavy neutrino consistent with the
correct baryon asymmetry via resonant leptogenesis. $\tan\beta$ is set to be $10$ in all cases. The values in the last two columns satisfy Eq.\eqref{eq:cons_Ib_dom} and Eq.\eqref{eq:cons_pdirac}. }
\end{table}
For each value of RH neutrino mass, three different values of coupling $Y_1$ are chosen around its extremum values when $R_1=0$ and its minimum value when $R_1=0.1$.

\begin{figure}[t!]
\begin{center}
\includegraphics[width=0.42\textwidth]{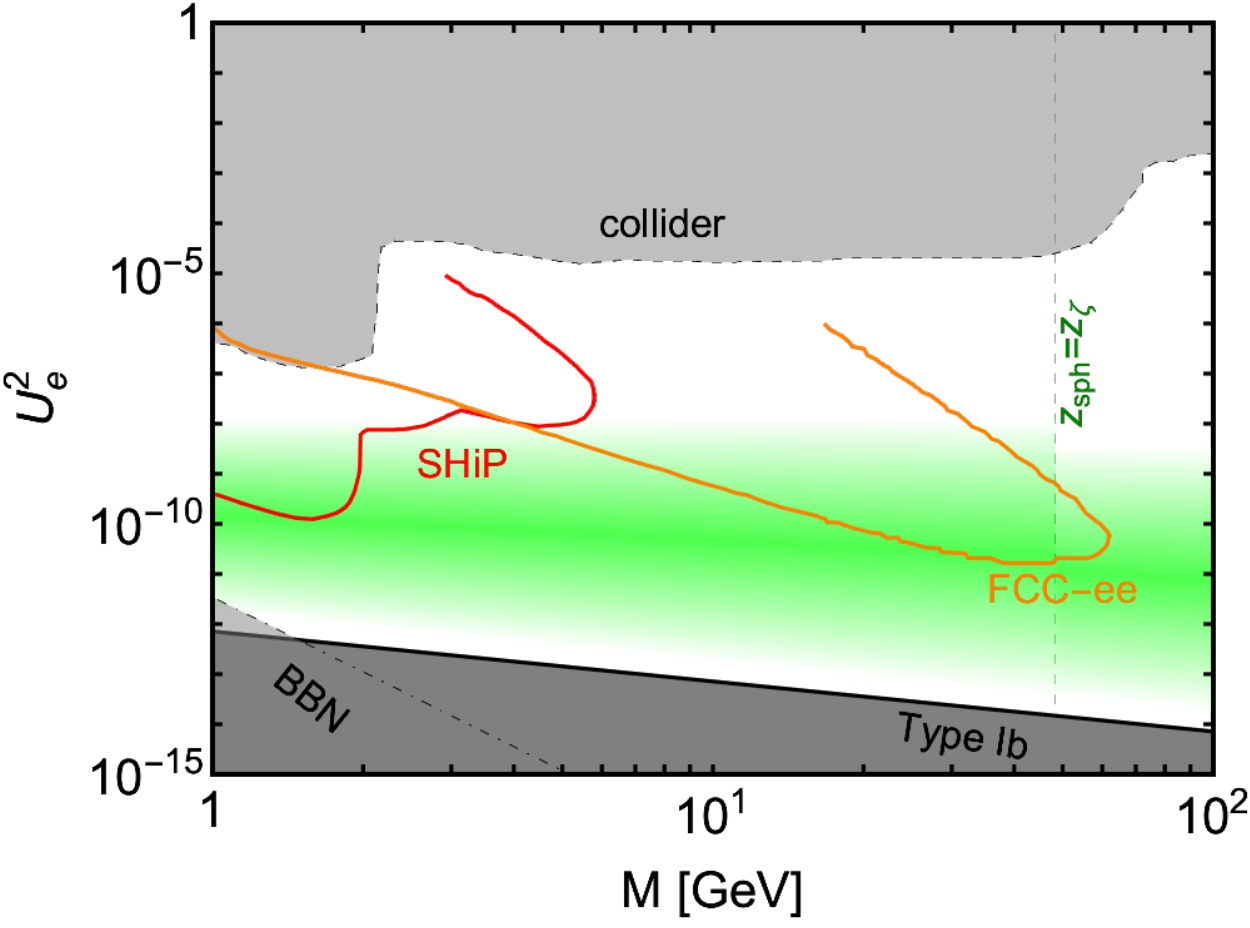}
\includegraphics[width=0.42\textwidth]{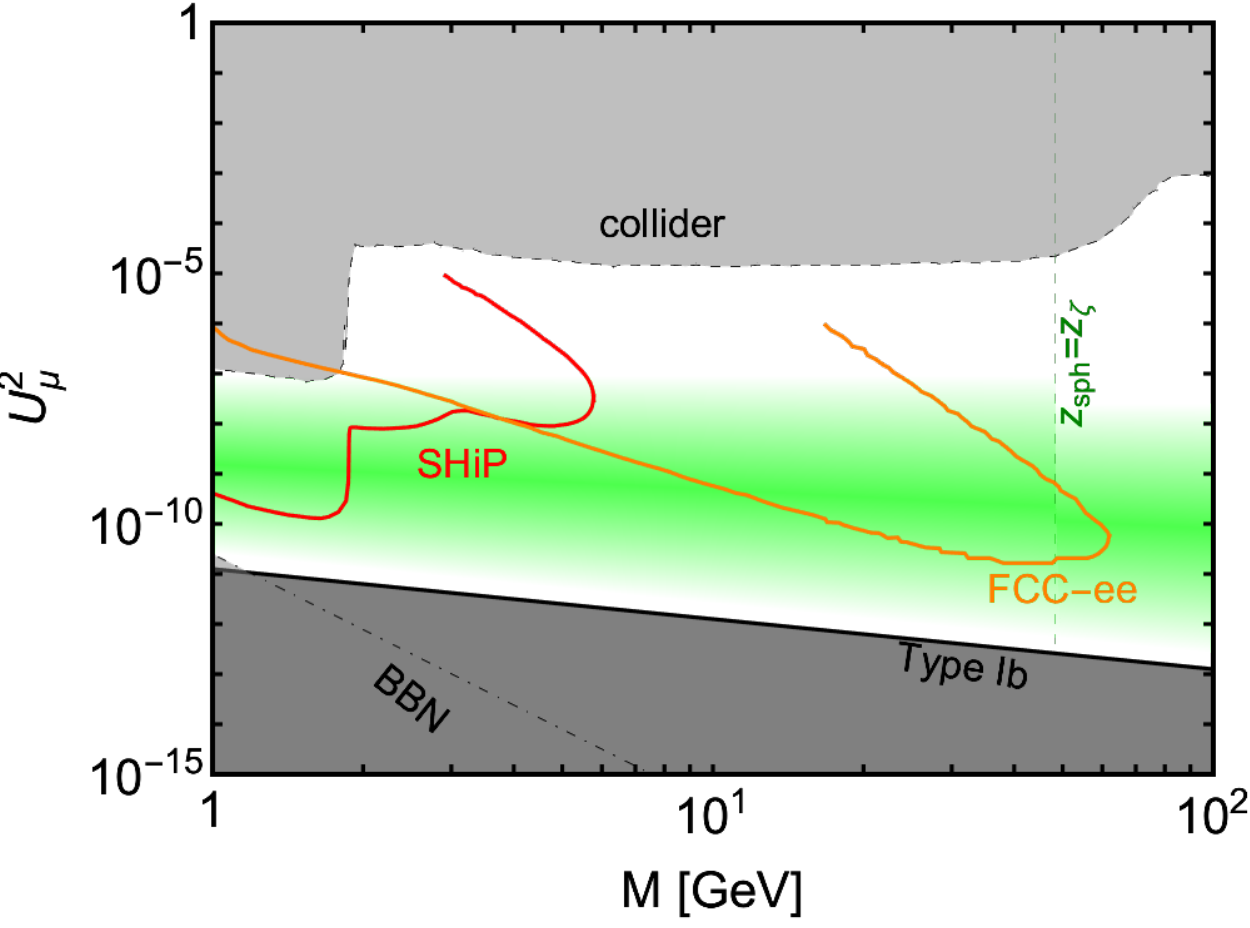}
\includegraphics[width=0.42\textwidth]{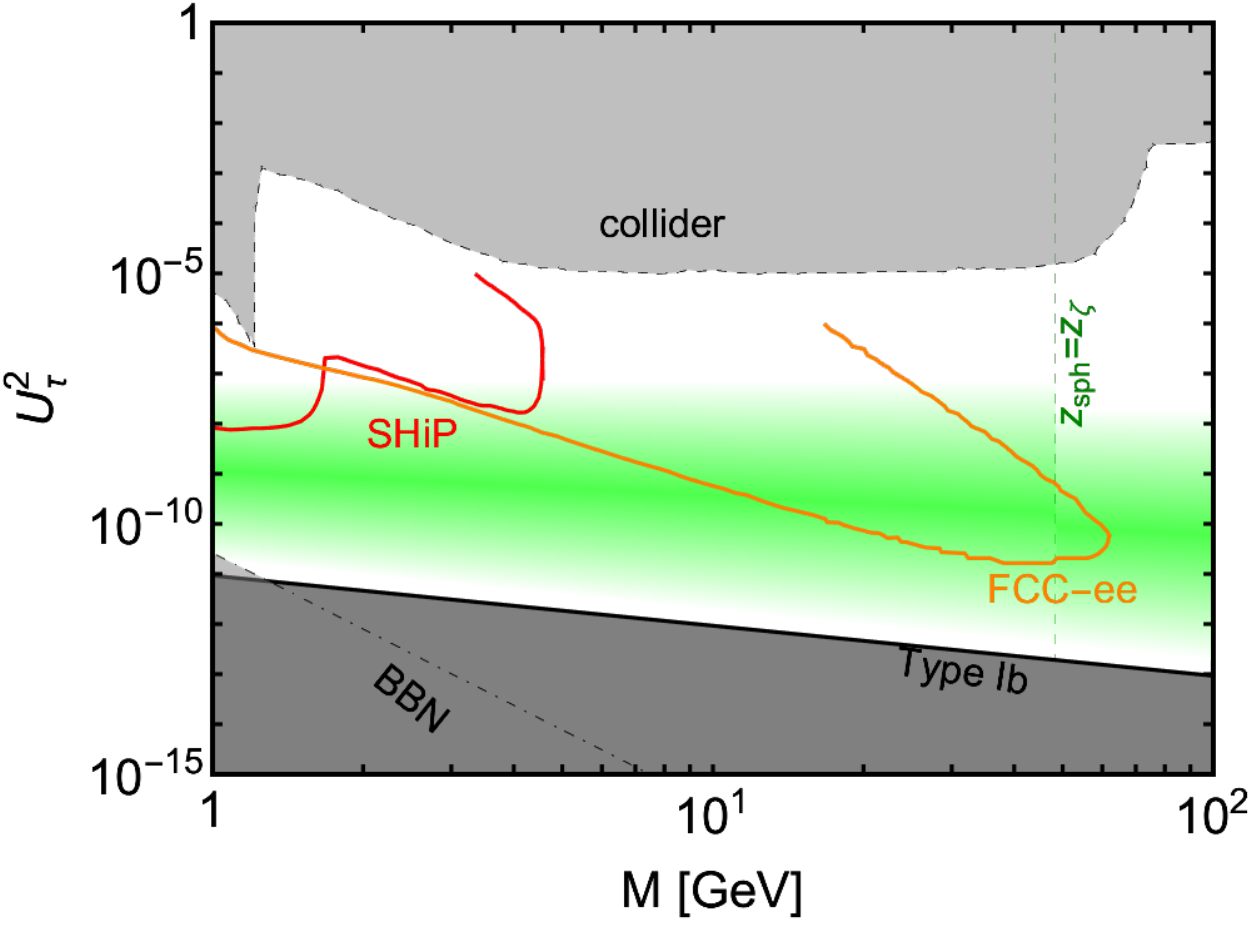}
\caption{\label{fig:U2} Allowed parameter space and constraints in the $M$-$U_\alpha^2$ plane, where $M$ is the Dirac mass of the right-handed neutrino pair in the type Ib seesaw model. The green region is the allowed parameter space for successful resonant leptogenesis in this model with transparent boundaries due to the strong inequalities in Eq.\eqref{eq2:U2}. See the text for detailed discussion. }
\end{center}
\end{figure}

As the constraints on the seesaw parameters from many existing and future collider experiments \cite{Drewes:2013gca,Blondel:2014bra,Drewes:2015vma,Drewes:2016jae,Deppisch:2015qwa,Chianese:2018agp,Beacham:2019nyx,SHiP:2018xqw,Drewes:2021nqr} involve the active-sterile neutrino mixing, we would like to derive the expression of active-sterile neutrino mixing strength in this model. The strength of the mixing between SM neutrinos and the heavy neutrino is represented by the quantity 
\begin{eqnarray}
U_\alpha^2=\sum_{i=1,2,3}|U_{\alpha i}|^2, \quad \alpha=e,\mu,\tau,
\label{eq1:U2}
\end{eqnarray}
where $U$ is the unitary matrix diagonalising the total neutrino mass matrix, including the SM model and RH neutrinos. The summation is over the index of RH neutrinos. More specifically, as $Y_1\gg Y_2$, the expressions of $U_\alpha^2$ for flavour $e$, $\mu$, $\tau$ are given by
\begin{subequations}
\label{eq2:U2}
\begin{align}
U_e^2&= (0.031+0.029\cos\delta_M) \frac{v_1^2 Y_1^2}{M^2},\\
U_\mu^2&= (0.27-0.16\cos\delta_M) \frac{v_1^2 Y_1^2}{M^2},\\
U_\tau^2&= (0.20+0.13\cos\delta_M) \frac{v_1^2 Y_1^2}{M^2},
\end{align}
\end{subequations}
where $\delta_M$ is an unconstrained relative Majorana phase in the PMNS mixing matrix. The allowed ranges of $U_\alpha^2$ are
\begin{subequations}
\label{eq2:U2}
\begin{align}
3.6 \times 10^{-7}\, \frac{\cos^2\beta}{\Theta(z_\zeta)} \gg U_e^2&\gg 2.2 \times 10^{-14} \frac{\cot\beta}{z_{sph}},\\
2.6 \times 10^{-6}\, \frac{\cos^2\beta}{\Theta(z_\zeta)} \gg U_\mu^2&\gg 1.3 \times 10^{-12} \frac{\cot\beta}{z_{sph}},\\
2.0 \times 10^{-6}\, \frac{\cos^2\beta}{\Theta(z_\zeta)} \gg U_\tau^2&\gg 8.0 \times 10^{-13} \frac{\cot\beta}{z_{sph}},
\end{align}
\end{subequations}
where the sphaleron decoupling temperature is taken to be $T_{sph}=131$ GeV \cite{DOnofrio:2014rug}. The allowed parameter space for resonant leptogenesis in the type Ib seesaw mechanism is shown as the green area in Fig.\ref{fig:U2} when $\zeta_0=23$ (the SM limit). The vertical green line marks out the threshold $z_{sph}=z_\zeta$. The other lines mark the constraints on the quantity $U_\alpha^2$ in the type Ib seesaw model. The parameter space is excluded by the neutrino data below the solid black line. The shadowed region above the dashed line is excluded by the collider data from multiple experiments~\cite{Deppisch:2015qwa} and the one below the dash-dotted line is excluded by Big Bang Nucleosynthesis (BBN) data~\cite{Canetti:2012kh,Drewes:2019mhg}. The red and orange lines indicate the future sensitivity of SHiP \cite{SHiP:2018xqw} and FCC-$ee$ \cite{Blondel:2014bra}. The allowed parameter space for resonant leptogenesis is not constrained by the current result from collider experiments, but is accessible to the future experiments SHiP and FCC-$ee$, allowing both the type Ib seesaw model and leptogenesis to be tested.

\subsection{The inclusion of dark matter}
So far we have considered leptogenesis in the type Ib seesaw model. Following the approach in \cite{Chianese:2021toe},
it is also possible to include dark matter (as well as leptogenesis) 
in the type Ib seesaw model
by adding a dark scalar $\phi$ and a dark Dirac fermion $\chi_{L,R}$, both odd under an extra dark matter symmetry $Z_2$, giving rise to the extra terms,
\begin{eqnarray}
&&\mathcal{L}_{\rm DS} = \overline{\chi}\left(i \slashed{\partial} - m_\chi \right)\chi + \left|\partial_\mu \phi\right|^2 - m^2_\phi \left|\phi\right|^2 + V\left(\phi\right), \quad\quad
\label{eq:lagDS} \\
&&\mathcal{L}_{\rm N_R portal} = y \phi \, (\overline{\chi_R}N^c_{R1} +  \overline{\chi_L}N_{R2}) + {\rm h.c.}.
\label{eq:ds}
\end{eqnarray}
These extra couplings will emerge in the low energy effective type Ib seesaw theory, 
providing that the dark scalar and the dark fermion have $Z_4$ charges $\phi \sim \omega$ and $\chi_{L,R}\sim \omega^2$
in the extended model, while being odd under the dark matter symmetry $Z_2$, with all other fields being even under $Z_2$. 
Dark matter is produced via the freeze-in mechanism via the RHN portal couplings, where the dark sector Yukawa coupling $y$ is typically required to be very small $y\lesssim 10^{-5}$ \cite{Chianese:2021toe}\footnote{The Higgs portal coupling is assumed to be negligible for simplicity in \cite{Chianese:2021toe}. Any non-negligible contribution can be treated separately and removed from the relic abundance when discussing the neutrino portal dark matter. The results would not be affected very much unless the dark matter is dominantly produced through the Higgs portal, which would correspond to the standard scenario which has been studied a lot in literature \cite{Silveira:1985rk,McDonald:1993ex,Burgess:2000yq,Patt:2006fw,Cline:2013gha,Athron:2017kgt,Chanda:2019xyl,Chianese:2020yjo}.}. Note that 
dark matter production is not affected by the third RH neutrino $N_{R3}$ and the singlet scalar $\xi$ as long as the scalar portal interaction is neglectable. 

With dark matter produced through the freeze-in mechanism, successful dark matter can be achieved over the entire allowed region of Fig.\ref{fig:U2} \cite{Chianese:2021toe}. As in \cite{Chianese:2021toe}, regions of dark matter may be delineated where the relic density depends on either the purely dark portal coupling $y$ or else a combination of $y$ and seesaw Yukawa couplings.
A lower bound on $U_\alpha^2$ can be derived above which the dark matter production can be determined partly by the seesaw Yukawa couplings.
The lower bound on $U_\alpha^2$ can be directly adopted from \cite{Chianese:2021toe}, and the results compared to 
Fig.\ref{fig:U2}, allowing a simultaneous test of the type Ib seesaw mechanism, leptogenesis and dark matter.


\section{Conclusion}

In this paper, we have studied three different models where the SM neutrino masses are generated through the type Ib seesaw mechanism. In the minimal type Ib seesaw model with 2RHNs and a single off-diagonal mass which can be regarded as a Dirac mass, 
the RH neutrino masses are completely degenerate and the lepton asymmetry can only be produced through the interference between the vertex diagram and tree-level diagram if the leptogenesis is flavour-dependent. In that case, we have found that the RH handed neutrino mass is very constrained above $10^{12}$ GeV, which is close to its upper bound for flavour-dependent leptogenesis and, although viable, does not allow a low scale type Ib seesaw model. 

In the type Ib seesaw model with a small mass splitting between the RH neutrinos, so that they become pseudo-Dirac,
we found that the asymmetry cannot be produced through resonant leptogenesis even if the resonant condition is satisfied. In fact, the asymmetry contributed by the wave-function diagram always vanishes as the Yukawa interaction of each Higgs doublet have only one universal phase in the couplings. More generally, we have shown that there is no contribution from wave-function diagram if a linear combination of the phases in the Yukawa couplings is $n\pi$ where $n$ is an integer. 

In order to circumvent these shortcomings, we have proposed a high energy extended type Ib seesaw model by including a third superheavy Majorana RH neutrino to supplement 
the low scale pseudo-Dirac pair. We also included a Higgs singlet whose VEV allows the pseudo-Dirac pair of neutrinos to mix with the superheavy RH neutrino. Below the third RH neutrino and Higgs VEV mass scales, 
an effective type Ib seesaw model emerges with a pseudo-Dirac mass splitting and a naturally small effective Yukawa coupling.
As the phase in the Yukawa couplings between the RH neutrinos and the Higgs singlet cannot be removed by redefining the neutrino fields, different phases appear in the seesaw Yukawa couplings of the type Ib seesaw Lagrangian, thereby making resonant leptogenesis viable. 

Within the resulting low energy effective type Ib seesaw model, we have determined the allowed range of the dominant seesaw couplings, consistent with successful resonant leptogenesis, where one of the Yukawa couplings is relatively large. We have presented a set of benchmark points for 1-100 GeV Dirac right-handed neutrino masses, where we
determined the allowed active-sterile neutrino mixing strength consistent with resonant leptogenesis.
We have shown that Dirac right-handed neutrinos in this mass range
are accessible to the future experiments SHiP and FCC-$ee$, allowing the type Ib seesaw mechanism with leptogenesis and dark matter to be tested.

\section*{Acknowledgments}
BF acknowledges the Chinese Scholarship Council (CSC) Grant No.\ 201809210011 under agreements [2018]3101 and [2019]536. SFK acknowledges the STFC Consolidated Grant ST/L000296/1 and the European Union’s Horizon 2020 Research and Innovation programme under Marie Sklodowska-Curie grant agreement HIDDeN European ITN project (H2020-MSCA-ITN-2019//860881-HIDDeN). The authors thank Marco Chianese for useful discussions.


\appendix
\section{Boltzmann Equations\label{ap:BE}}

In general, the contribution to the lepton number asymmetry $\Delta L_{\alpha}$ can be parted into a source term and a washout term 
\begin{eqnarray}
\frac{dY_{\Delta L_{\alpha}}}{dz} &=& \left(\frac{dY_{\Delta L_{\alpha}}}{dz}\right)^s + \left(\frac{dY_{\Delta L_{\alpha}}}{dz}\right)^w \label{eq:BE_LNA}.
\end{eqnarray} 
In the type Ib seesaw model, the expression of the lepton asymmetry is slightly different from in the troditional seesaw model due to the distinction between the two Higgs doublets. In general, it can be shown that the CP aymmetry in scattering processes is the same as in decays and inverse decays by considering the phase space integral \cite{Abada:2006ea,Davidson:2008bu}. However, as the top quark only couples to the second Higgs doublet $\phi_2$, only $\phi_2$ can play the roll of intermediate on-shell particle in the top quark scattering. Therefore the CP aymmetry in top quark scattering is only contributed by the second Higgs doublet $\phi_2$, i.e. $\Delta\gamma^{n_i\overline{t}}_{\ell_\alpha\overline{q}_3} \simeq \gamma^{n_i\overline{t}}_{\ell_\alpha\overline{q}_3}\left(\epsilon_{n_i}\right)_{2\alpha}$, while the CP aymmetry in guage boson scattering recieves contribution from both Higgs doublets $\Delta\gamma^{n_i A}_{\ell_\alpha\overline{\phi}_k} \simeq \sum_l\gamma^{n_i A}_{\ell_\alpha\overline{\phi}_l}\left(\epsilon_{n_i}\right)_{k\alpha}$. After substition, the source term reads
\begin{eqnarray}
&& \left(\frac{dY_{\Delta L_{\alpha}}}{dz}\right)^s \simeq \frac{z}{\s\Hub(M)} \sum_{i}\left(\frac{Y_{n_i}}{Y^{\rm eq}_{n_i}} -1\right) \times \nonumber\\
&&\quad\quad \left[ \sum_{k}\left(\gamma_{n_i\rightarrow2} + \gamma_{2\rightarrow2}^{n_iA} \right) \left(\epsilon_{n_i}\right)_{k\alpha} + \gamma_{2\rightarrow2}^{n_it} \left(\epsilon_{n_i}\right)_{2\alpha}\right].\quad\quad
\end{eqnarray}
The washout term in Eq.\eqref{eq:BE_LNA} reads 
\begin{widetext}
\begin{eqnarray}
\left(\frac{dY_{\Delta_{\alpha}}}{dz}\right)^w &=& \sum_{k,l,\beta}\left[ (\Delta r_{Y_{\ell_\alpha}} +\Delta r_{Y_{\phi_k}})\left(\gamma^{\ell_\alpha\phi_k}_{\overline{\ell}_\beta\overline{\phi}_l} + \gamma^{\ell_\alpha\phi_k}_{\ell_\beta\phi_l}\right) + (\Delta r_{Y_{\ell_\beta}} +\Delta r_{Y_{\phi_k}})\left(\gamma^{\ell_\alpha\phi_k}_{\overline{\ell}_\beta\overline{\phi}_l} - \gamma^{\ell_\alpha\phi_k}_{\ell_\beta\phi_l}\right) \right]\nonumber\\
&+&\sum_{k,l,\beta}\left[ (1+\delta_{\alpha\beta})(\Delta r_{Y_{\ell_\alpha}} + \Delta r_{Y_{\ell_\beta}} + \Delta r_{Y_{\phi_k}} + \Delta r_{Y_{\phi_l}})\gamma^{\ell_\alpha\ell_\beta}_{\overline{\phi}_k\overline{\phi}_l} + (\Delta r_{Y_{\ell_\alpha}} - \Delta r_{Y_{\ell_\beta}}) \gamma^{\ell_\alpha\overline{\ell}_\beta}_{\overline{\phi}_k\phi_k} \right]\nonumber\\
&+& \sum_i \left[(r_{Y_{n_i}}\Delta r_{Y_{\ell_\alpha}} - \Delta r_{Y_{q_3}} + \Delta r_{Y_{t}}) \gamma^{n_i\ell_\alpha}_{q_3\overline{t}} + \left[2\Delta r_{Y_{\ell_\alpha}} - (r_{Y_{n_i}}+1)(\Delta r_{Y_{q_3}} - \Delta r_{Y_{t}}) \right]\gamma^{n_iq_3}_{t\ell_\alpha}\right]\nonumber\\
&-&\sum_{i,k}\left[(r_{Y_{n_i}}\Delta r_{Y_{\ell_\alpha}} +\Delta r_{Y_{\phi_k}})\gamma^{n_i\ell_\alpha}_{A\overline{\phi}_k} + (r_{Y_{n_i}}\Delta r_{Y_{\phi_k}}+\Delta r_{Y_{\ell_\alpha}} )\gamma^{n_i\phi_k}_{A\overline{\ell}_\alpha} + (\Delta r_{Y_{\phi_k}}+\Delta r_{Y_{\ell_\alpha}} )\gamma^{n_iA}_{\phi_k\ell_\alpha}\right]\,,\label{eq:washout}
\end{eqnarray} 
\end{widetext} 
where $\Delta r_{Y_a}=(Y_a-Y_{\overline{a}})/Y_a^{\rm eq}$. The asymmetries in various types of particles are involved, which can be connected to the asymmetries in lepton doublets through spectator processes as discussed in Sec.\ref{ap:Amatrix} \cite{Nardi:2006fx}. Finally, the asymmetry in lepton doublets can be related to the $B-L_\alpha$ asymmetry by the $A$-matrix ad the washout term reduces into
\begin{eqnarray}
\left(\frac{dY_{\Delta L_{\alpha}}}{dz}\right)^w &=&-\frac{z}{\s\Hub(M)}\sum_\beta \frac{A_{\alpha\beta}Y_{\Delta_{\beta}}}{Y_\ell^{\rm eq}} \\ && \left( \gamma_{W,1\rightarrow2}^{n_i} + \gamma_{W,2\rightarrow2}^{n_i t} + \gamma_{W,2\rightarrow2}^{n_iA} \right)_\alpha\,,\quad\quad
\end{eqnarray} 
where $Y_\ell^{\rm eq}$ is the equalibrium comoving density of leptons and $\gamma_W$ denotes the effective cross section from decay and scattering processes contributing to the washout, whose expression can be read from Eq.\eqref{eq:washout}.

\section{``$A$''-matrix\label{ap:Amatrix}}
The aim of this appendix is to show how the ``$A$''-matrix in the minimal type Ib seesaw leptogenesis is computed explicitly. The asymmetries in the particle number densities $n_X$ is determined by the temperature and chemical potential through \cite{Kolb:1990vq}
\begin{eqnarray}
n_X-\overline{n}_X = 
\begin{dcases}
\frac{g_X T^2}{6} \mu_X& \text{fermions,}\\
\frac{g_X T^2}{3} \mu_X& \text{bosons.}
\end{dcases} 
\end{eqnarray} 
If an interaction is in equilibrium, the sum of the chemical potentials over all particles entering the interaction should vanish. Different kinds of interactions can be in equilibrium at the temperatures that the lepton asymmetry is produced, therefore the relation between the chemical potentials can vary. These interactions in equilibrium are also referred to as the ``spectator processes'' \cite{Davidson:2008bu}. In the flavour-dependent leptogenesis, the $B-L_\alpha$ asymmetry takes the expression 
\begin{eqnarray}
Y_{\Delta_{\alpha}} &=& \left[\sum_{\beta=1,2,3}\frac{2\mu_{q_\beta}+\mu_{{u_R}_\beta}+\mu_{{d_R}_\beta}}{3} - \left(2\mu_{\ell_\alpha}+\mu_{{e_R}_\alpha}\right) \right] \nonumber\\ &&\times\frac{T^2}{6s}.
\label{ap_eq:B-L}
\end{eqnarray} 

In the minimal type Ib seesaw model, assuming that the gauge multiplet have the same chemical potential and the gauge bosons have zero chemical potentials, there are 17 independent chemical potentials: 6 for SM fermion doublets, 9 for SM fermion singlets and 2 for Higgs doublets. In the temperature range $\left\{(1+\tan^2\beta)10^9,\,(1+\tan^2\beta)10^{12}\right\}$ GeV, there are following relations between the chemical potentials \cite{Nardi:2006fx}: 
\begin{itemize}
\item{the strong sphaleron} 
\begin{eqnarray}
\sum_{\beta=1,2,3}\left(2\mu_{q_\beta}-\mu_{{u_R}_\beta}-\mu_{{d_R}_\beta} \right)&=&0.
\end{eqnarray} 
\item{the electroweak sphaleron}  
\begin{eqnarray}
3\sum_{\beta=1,2,3}\mu_{q_\beta} + \sum_{\alpha=e,\mu,\tau}\mu_{\ell_\beta}&=&0.
\end{eqnarray} 
\item{the equilibrium of top, bottom, charm and tau Yukawa interactions}
\begin{eqnarray}
\mu_{t_R}-\mu_{q_3}+\mu_{\phi_2}&=&0,\\
\mu_{b_R}-\mu_{q_3}-\mu_{\phi_1}&=&0,\\
\mu_{c_R}-\mu_{q_2}+\mu_{\phi_2}&=&0,\\
\mu_{\tau_R}-\mu_{\ell_\tau}-\mu_{\phi_1}&=&0.
\end{eqnarray} 
\item{the equality of asymmetries in different baryon flavours $Y_{\Delta B_1}=Y_{\Delta B_2}=Y_{\Delta B_3}$ }
\begin{eqnarray}
2\mu_{q_1}+\mu_{u_R}+\mu_{d_R}&=&2\mu_{q_2}+\mu_{c_R}+\mu_{s_R}\nonumber\\ &=&2\mu_{q_3}+\mu_{t_R}+\mu_{b_R}.
\end{eqnarray} 
\item{the hypercharge neutrality}
\begin{eqnarray}
&&\sum_{\beta=1,2,3}\left(\mu_{q_\beta} + 2\mu_{{u_R}_\beta}-\mu_{{d_R}_\beta} \right) \nonumber\\ &&- \sum_{\alpha=e,\mu,\tau}\left(\mu_{\ell_\beta} +\mu_{e_\beta}\right) -2\mu_{\phi_1}-2\mu_{\phi_2}=0.
\end{eqnarray} 
\item{the neglectable asymmetries in $\mu_R$, $e_R$, $u_R-d_R$ and $d_R-s_R$}
\begin{eqnarray}
\mu_{\mu_R}&=&0,\label{ap_eq:muon}\\
\mu_{e_R}&=&0,\label{ap_eq:electron}\\
\mu_{u_R}-\mu_{d_R}&=&0,\label{ap_eq:u-d}\\
\mu_{d_R}-\mu_{s_R}&=&0\,\label{ap_eq:d-s}.
\end{eqnarray} 
\item{the equilibrium of the interaction between Higgs doublets}
\begin{eqnarray}
\mu_{\phi_1}=\mu_{\phi_2}.
\end{eqnarray} 
\end{itemize}
In total, there are 14 equalities between the chemical potentials, therefore it is possible to express the Eq.\eqref{ap_eq:B-L} with the chemical potential of three lepton doublets as
\begin{eqnarray}
&&\begin{pmatrix}
Y_{\Delta_{e}} \\
Y_{\Delta_{\mu}} \\
Y_{\Delta_{\tau}} 
\end{pmatrix}
=
\frac{T^2}{6s} 
\begin{pmatrix}
-22/9 & -4/9 & -4/9 \\
-4/9 & -22/9 & -4/9 \\
-46/171 & -46/171 & -541/171
\end{pmatrix}
\begin{pmatrix}
\mu_{\ell_e} \\
\mu_{\ell_\mu} \\
\mu_{\ell_\tau} 
\end{pmatrix}
\nonumber\\&&=
\begin{pmatrix}
-11/9 & -2/9 & -2/9 \\
-2/9 & -11/9 & -2/9 \\
-23/171 & -23/171 & -541/342
\end{pmatrix}
\begin{pmatrix}
Y_{\Delta_{\ell_e}} \\ 
Y_{\Delta_{\ell_\mu}} \\
Y_{\Delta_{\ell_\tau}}
\end{pmatrix}.
\end{eqnarray}
However, as the lepton flavours $e$ and $\mu$ are not distinguishable, the chemical potential $\mu_{\ell_e}$ and $\mu_{\ell_\mu}$ should be summed
\begin{eqnarray}
\begin{pmatrix}
Y_{\Delta_{e+\mu}} \\
Y_{\Delta_{\tau}} 
\end{pmatrix}
&= &
\begin{pmatrix}
-13/9 & -4/9 \\
-23/171 & -541/342
\end{pmatrix}
\begin{pmatrix}
Y_{\Delta_{\ell_{e+\mu}}} \\
Y_{\Delta_{\ell_\tau}} 
\end{pmatrix} .\quad\quad
\end{eqnarray}
The ``$A$''-matrix is defined by the linear transformation from $Y_{\Delta_{\alpha}}$ to $Y_{\Delta_{\ell_\alpha}}$ and therefore 
\begin{eqnarray}
A=\begin{pmatrix}
-541/761 & 152/761\\
46/761 & -494/761
\end{pmatrix}.
\end{eqnarray}

At temperature between $(1+\tan\beta^2)\times10^5$ and $(1+\tan\beta^2)\times10^9$ GeV, the strange quark and muon Yukawa interactions are also in equilibrium. Therefore Eq.\eqref{ap_eq:muon} and Eq.\eqref{ap_eq:d-s} are substitute by 
\begin{eqnarray}
\mu_{s_R}-\mu_{q_2}-\mu_{\phi_1}&=&0,\\
\mu_{\mu_R}-\mu_{\ell_\mu}-\mu_{\phi_1}&=&0.
\end{eqnarray} 
The $Y_{\Delta_{\alpha}}-Y_{\Delta_{\ell_\alpha}}$ relation becomes
\begin{eqnarray}
\begin{pmatrix}
Y_{\Delta_{e}} \\
Y_{\Delta_{\mu}} \\
Y_{\Delta_{\tau}} 
\end{pmatrix}
=
\begin{pmatrix}
-11/9 & -2/9 & -2/9 \\
-11/72 & -29/18 & -1/9 \\
-11/72 & -1/9 & -29/18
\end{pmatrix}
\begin{pmatrix}
Y_{\Delta_{\ell_e}} \\ 
Y_{\Delta_{\ell_\mu}} \\
Y_{\Delta_{\ell_\tau}} 
\end{pmatrix}\quad\quad
\end{eqnarray}
and the ``$A$''-matrix is
\begin{eqnarray}
A=\begin{pmatrix}
-93/110 & 6/55 & 6/55\\
3/40 & -19/30 & 1/30\\
3/40 & 1/30 & -19/30
\end{pmatrix}.
\end{eqnarray}

At temperature below $(1+\tan\beta^2)\times10^5$ and before the electroweak sphaleron decouples, all the SM Yukawa interactions are in equilibrium. In that case, the QCD sphaleron condition becomes redundant and can be ignored. Besides, Eq.\eqref{ap_eq:electron} and Eq.\eqref{ap_eq:u-d} no longer hold as the asymmetries in $e_R$ and $u_R-d_R$ are not conserved. Instead, there are 
\begin{eqnarray}
\mu_{u_R}-\mu_{q_1}+\mu_{\phi_2}=0, &&\quad \mu_{d_R}-\mu_{q_1}-\mu_{\phi_1}=0 ,\nonumber\\
\mu_{e_R}-\mu_{\ell_e} &-& \mu_{\phi_1}=0.
\end{eqnarray}
The ``$A$''-matrix is then
\begin{eqnarray}
A&=&\begin{pmatrix}
-128/207 & 10/207 & 10/207\\
10/207 & -128/207 & 10/207\\
10/207 & 10/207 & -128/207
\end{pmatrix}.
\end{eqnarray}

\bibliography{Type1b}

\end{document}